\documentclass[aps,pre,preprint,onecolumn]{revtex4-1} 	

\usepackage{graphicx, amsmath, amssymb, lineno}
\usepackage{hyperref}
\hypersetup{ 
	colorlinks   = true
}
\usepackage{color,comment}
\usepackage[x11names]{xcolor}
\newif\ifproofread

\begin{abstract}
Surveys of microbial biodiversity such as the Earth Microbiome Project (EMP) and the Human Microbiome Project (HMP) have revealed robust ecological patterns across different environments. A major goal in ecology is to leverage these patterns to identify the ecological processes shaping microbial ecosystems. One promising approach is to use minimal models that can relate mechanistic assumptions at the microbe scale to community-level patterns. Here, we demonstrate the utility of this approach by showing that the Microbial Consumer Resource Model (MiCRM) -- a minimal model for microbial communities with resource competition, metabolic crossfeeding and stochastic colonization --  can qualitatively reproduce patterns found in survey data including compositional gradients, dissimilarity/overlap correlations, richness/harshness correlations, and nestedness of community composition. By using the MiCRM to generate synthetic data with different environmental and taxonomical structure, we show that large scale patterns in the EMP  can be reproduced by considering the energetic cost of surviving in harsh environments and HMP patterns may reflect the importance of environmental filtering in shaping competition. We also show that recently discovered dissimilarity-overlap correlations in the HMP likely arise from communities that share similar environments rather than reflecting universal dynamics. We identify ecologically meaningful changes in parameters that alter or destroy each one of these patterns, suggesting new mechanistic hypotheses for further investigation. These findings highlight the promise of minimal models for microbial ecology.
\end{abstract}
\begin{document}
\title{A minimal model for microbial biodiversity can reproduce experimentally observed ecological patterns}
\author{Robert Marsland III}
\email{marsland@bu.edu}
\author{Wenping Cui} 
\altaffiliation{Department of Physics, Boston College, Chestnut Hill, MA}
\author{Pankaj Mehta}
\affiliation{Department of Physics, Boston University, Boston, MA}

\date{\today}

\maketitle

Over the past decade, next-generation sequencing has highlighted the incredible diversity of the microbial ecosystems that fill every corner of our planet.  Microbial communities are incredibly complex and and occur in environments ranging from soils to the human body. Large-scale surveys of microbial biodiversity, such as the Earth Microbiome Project (EMP), the Human Microbiome Project (HMP) and the European Metagenomics of the Human Intestinal Tract project (MetaHIT), have revealed a number of robust and reproducible patterns in community composition and function \cite{EMP,HMP,qin2010human}. A major challenge for contemporary microbial ecology is to understand and identify the ecological origins of these patterns. This problem is especially difficult because it involves what in the ecology literature has been called the ``problem of pattern and scale'' \cite{levin1992problem}: explaining ecological patterns requires connecting processes that occur at very different scales of spatial, temporal, and taxonomical organization. 

One potential approach for overcoming the problem of scale is to use mathematical models and large simulations to investigate how mechanistic assumptions about environmental and taxonomical structure at the microscopic scale affect the kind of  ecological patterns observed at larger scales. A major obstacle in realizing this goal is that any mathematical model that seeks to explain modern microbial sequencing data must deal with the enormous complexity of microbial communities: the numbers of species and consumable molecules in a community can easily reach into the hundreds or thousands \cite{EMP}. Thus, by necessity any mechanistic model of community assembly will have an extraordinary number of free parameters, presenting a major obstacle for understanding microbial dynamics \cite{hart2019uncovering}.

\begin{figure*}
	\centering
	\includegraphics[width=17cm]{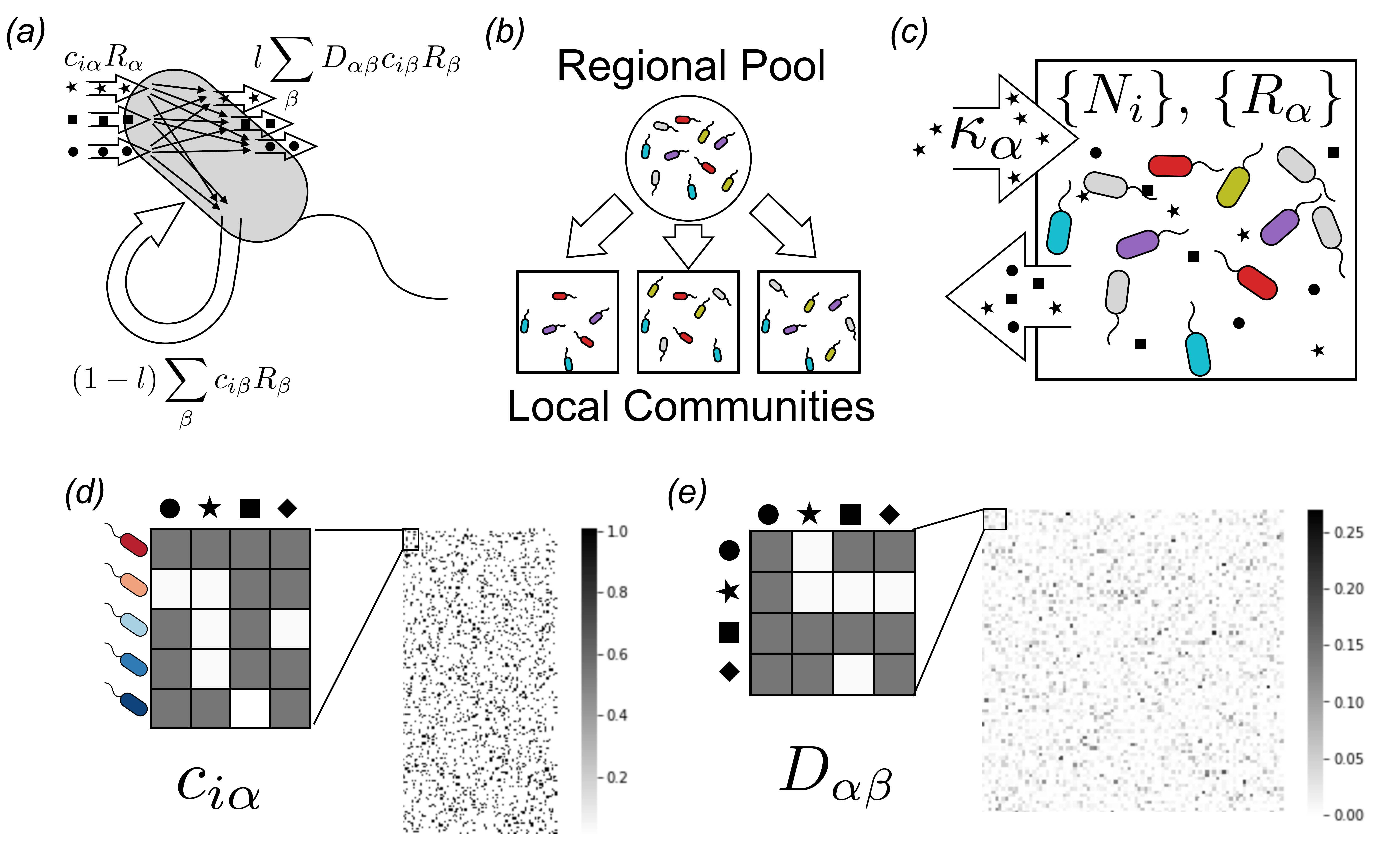}
	\linespread{1}\selectfont{}
	\caption{{\bf A minimal model for investigating microbial biodiversity.} \emph{(a)} The Microbial Consumer Resource Model extends the classic Consumer Resource Model of MacArthur and Levins \cite{MacArthur1970} by incorporating the generic exchange of secondary metabolites observed in microbial communities, as described in the Methods. Each consumed resource type $\alpha$ (stars, squares, circles) with abundance $R_\alpha$ is taken up by species $i$ at a rate $c_{i\alpha}R_\alpha$, and transformed into other resource types through metabolic reactions inside each cell with normalized stoichiometry matrix $D_{\alpha\beta}$. A fraction $l$ of the resulting chemical flux returns to the environment, where it can be consumed by other microbes, while the rest is retained and used for growth. \emph{(b)} Communities are initialized by randomly sampling subsets of species from a given regional pool, simulating the effect of stochastic colonization. The importance of dispersal limitation for community assembly can be tuned by adjusting the number of species in each the subset. \emph{(c)} Each community is supplied with a constant influx of specified resource types, and all resources are diluted at a fixed rate. We assume that each community is well-mixed, so that its state is fully defined by the set of resource abundances $R_\alpha$ and microbial population sizes $N_i$. \emph{(d)} Heat map of randomly sampled matrix of consumer preferences $c_{i\alpha}$ with $S=200$ species and $M=100$ resource types. \emph{(e)} Heat map of randomly sampled metabolic matrix $D_{\alpha\beta}$, which encodes the allowed metabolic transformations and their relative rates, shown here with $M=100$ resource types.}
	\label{fig:model}
\end{figure*}

One potential strategy for overcoming this difficulty is  to exploit the observation that complex systems often have generic behaviors that can be described by sampling parameters from an appropriately chosen random distribution \cite{Wigner1955, May1972}.  The most famous example of this is in nuclear physics where the intractably complicated quantum dynamics of the uranium nucleus were successfully modeled using random matrices \cite{Wigner1955}.  Recently, we have adapted these ideas to the microbial setting by formulating a minimal model for microbial population dynamics we term the Microbial Consumer Resource Model (MiCRM) \cite{Goldford2018, marsland2018available, marsland2019community} (see Figure \ref{fig:model}). 

The MiCRM builds on the classic framework for resource competition developed by MacArthur and Levins\cite{MacArthur1970}. As in all consumer resource models,  species in the MiCRM are defined by their preferences for resources (Fig. \ref{fig:model}d). Species with similar preferences naturally compete with each other, giving rise to competitive exclusion and niche partitioning. Crucially, the MiCRM incorporates two additional pieces of biological knowledge that are specific to microbial communities. First, the MiCRM explicitly includes cross-feeding and syntrophy -- the consumption of metabolic  byproducts  of  one  species  by  another  species \cite{Harcombe2014, Zomorrodi2016,Goldford2018,pacheco2019costless}. This is incorporated into the MiCRM through a stoichiometric metabolic matrix that parameterizes the metabolic transformations of consumed metabolites into secreted byproducts (Fig. \ref{fig:model}a,e).  Second the MiCRM incorporates stochasticity in dispersal and colonization\cite{leibold2004metacommunity,vellend2016theory,hillerislambers2012rethinking,dini2018embracing}. Due to proximity effects, it is known that new environments are almost always colonized by only a subset of all the species capable of existing in that environment \cite{shurin2000dispersal}.  The MiCRM incorporates stochastic dispersal by seeding new environments through random sampling of a larger regional species pool (Fig. \ref{fig:model}b).

Taxonomic and metabolic assumptions are incorporated into the MiCRM through the choice of consumer preferences and metabolic matrices (see Methods and \cite{marsland2019community} for detailed discussion and implementation details). In the most minimal version of the MiCRM, species have no taxonomic structure (i.e. consumer preferences are uncorrelated across species and resources) and metabolism is completely random (i.e. the metabolic matrix has no structure beyond that required by energy and mass conservation). Large-scale surveys such as EMP and HMP often sample communities from very different environmental conditions. For this reason, it is important to be able to incorporate environmental structure and heterogeneity into our models. This is done by choosing which externally supplied resources are present in an environment (Fig. \ref{fig:model}c).

Importantly,  the MiCRM  also allows for the incorporation of additional metabolic and taxonomic structure allowing us to ask how taxonomy and metabolism shape community structure and function. This is implemented in the MiCRM by dividing resources into general resource classes (e.g. sugars, carboxylic acids, lipids, amino acids, etc.) and then using a tiered secretion model where metabolic resources are preferentially secreted into certain resource classes (Methods and \cite{marsland2019community}). This allows us to incorporate metabolic structure missing in the minimal MiCRM such as the fact that the fermentation of sugars  preferentially results in the secretion of carboyxlic acids. 

Taxonomic structure can also be easily incorporated into the MiCRM by introducing correlations in species preferences that come from the same family. For example, it is well known that bacteria from the \emph{Enterobacteria} family have a strong preference for fermenting sugars. The MiCRM incorporates such preferences by assigning species to families, with each family preferentially consuming resources from certain resource classes. Importantly, we can control the amount of metabolic and taxonomic structure in the community by modulating just two parameters that control the correlation structures of the consumer preference and metabolic matrix  (see Methods and \cite{marsland2019community}). This  addresses the major modeling bottleneck discussed above about how to choose parameters for diverse ecosystems.

In this paper, we use the MiCRM to test simple hypotheses about the mechanistic origins of patterns observed in EMP, HMP and MetaHIT, as well as in recent studies of marine microbial communities \cite{Sunagawa2015,enke2018modular}. We find that the MiCRM can qualitatively reproduce observed phenomena with minimum fitting or fine-tuning. We illustrate the utility of the model by identifying ecological mechanisms necessary for reproducing observed patterns as well as identifying ecological processes that can destroy these patterns. This allows us to use the MiCRM to generate new ecological hypotheses linking microscopic processes to large-scale patterns

All simulation data and analysis scripts are available at\\ \url{https://github.com/Emergent-Behaviors-in-Biology/microbiome-patterns}.\\ The model itself is implemented in the freely available Python module Community-Simulator \cite{marsland2019community} \url{https://github.com/Emergent-Behaviors-in-Biology/community-simulator}.\\ Since the number of simulations required for comparisons with survey data is necessarily large, our numerical work relies heavily on a novel algorithm implemented in the Community Simulator, which takes advantage of a recently discovered duality between consumer resource models and constrained optimization to quickly and accurately simulate hundreds of communities\cite{mehta2018constrained,Marsland2019a}. 

\section*{Results}

\begin{figure*}[t]
	\centering
	\includegraphics[width=16cm]{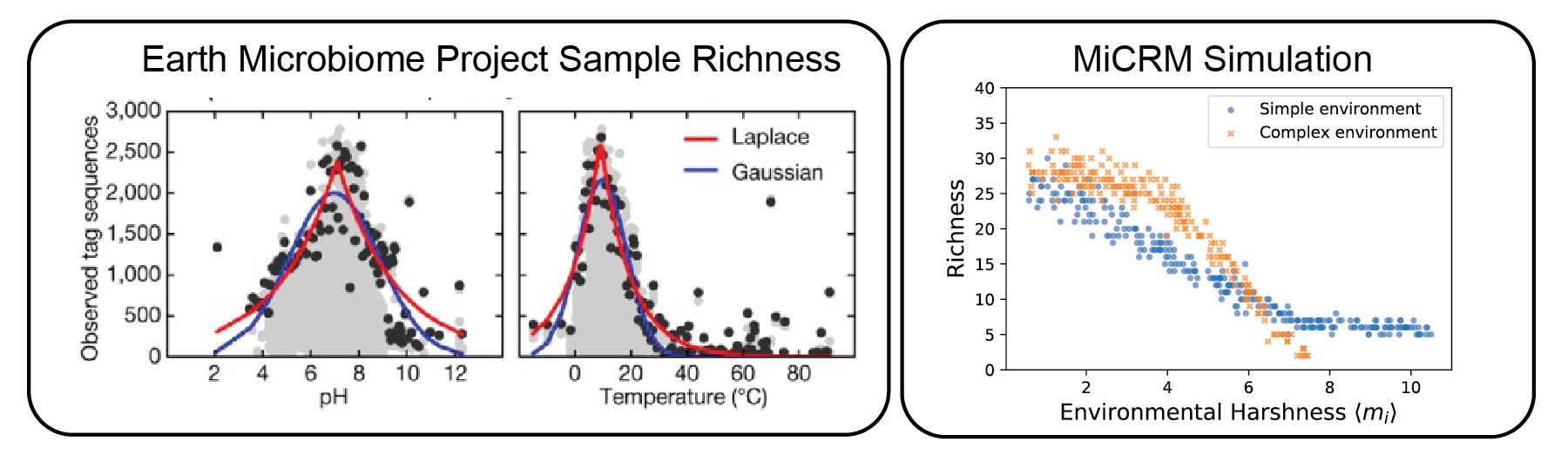}
	\linespread{1}\selectfont{}
	\caption{{\bf Relationship between diversity and environmental harshness is modulated by environmental complexity.} Left: Gray dots are the number of distinguishable strains observed in each sample of the EMP, plotted vs. pH and temperature. Black dots represent the 99th percentile of all communities at a given pH or temperature. Colored lines are fits of a Laplacian and a Gaussian distribution to the 99the percentile points. Reproduced from Figure 2 of the initial open-access report on the results of the EMP\cite{EMP}.  Right: The number of species surviving to steady state in simulated communities, plotted vs. environmental harshness. Harsher environments at extreme pH or temperature were simulated by increasing the total amount of resource consumption $m_i$ required for growth (by the same amount for all species). Blue squares are simulation results when all the energy was supplied via a single resource type, while orange circles are simulations where the incoming energy was evenly divided over all 90 possible resource types. See main text and Methods for simulation details.}
	\label{fig:EMP}
\end{figure*}

\begin{figure*}[t]
	\centering
	\includegraphics[width=16cm]{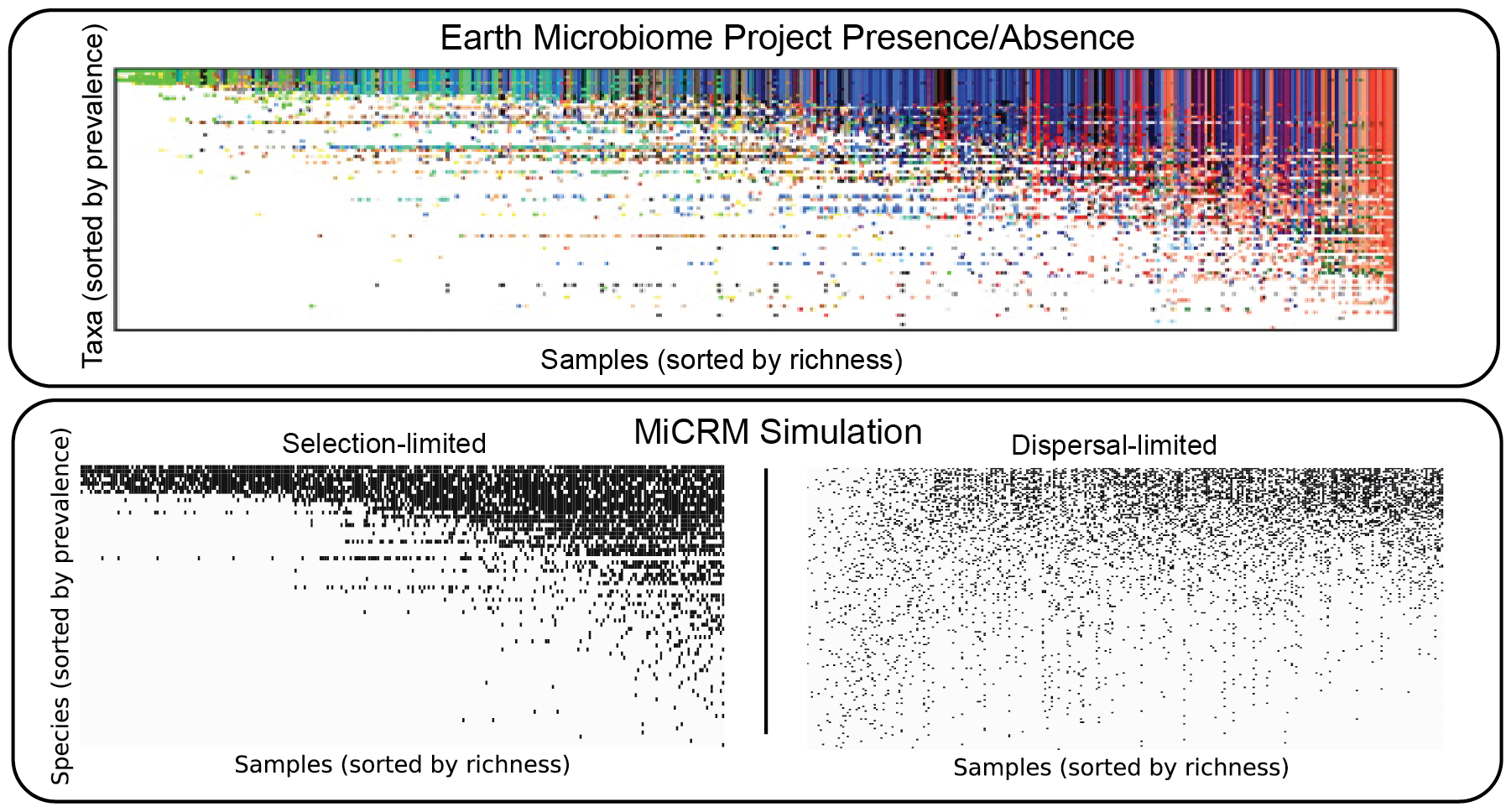}
	\linespread{1}\selectfont{}
	\caption{{\bf Nestedness of community composition indicates selection-dominated community assembly.} Top: Presence (colored) or absence (white) of each microbial phylum in a representative set of 2,000 samples from the EMP. Reproduced from Figure 3 of the EMP report \cite{EMP}. Different colors represent different biomes. Bottom: Presence (black) or absence (white) of species in simulated communities. Two different regimes of community assembly were simulated. The first is the selection-dominated scenario of Figure \ref{fig:EMP}, where variability in diversity is produced by variations in environmental harshness, and all samples are initialized with the vast majority (150/180) of the species in the regional pool. The second is a dispersal-dominated scenario, where environmental conditions are identical for all samples, but each sample is initialized with a different number of species, varying from 1 to 180. See main text and Methods for simulation details.}
	\label{fig:nest}
\end{figure*}

\subsection*{Patterns in the Earth Microbiome Project can be explained by energetic costs associated with harsh environments}

The Earth Microbiome Project is a systematic attempt to characterize  global microbial diversity and function. It consists of over 20,000 samples in 17 environments located on all 7 continents \cite{EMP}. Recently, a metaanalysis of this data was carried out and several robust patterns were identified. Chief among these was an interesting anti-correlation between richness and environmental harshness reproduced in Figure \ref{fig:EMP}. Samples near neutral pH or at moderate temperatures ($\sim 15^\circ$C) showed much higher levels of richness than samples from more extreme conditions. Peak richness dropped by a factor of 2 for pHs less than 5 or greater than 9, and temperatures less than 5 $^\circ$C or greater than 20 $^\circ$C. 

The EMP samples also showed a strongly nested structure: less diverse communities tended to be subsets of the more diverse communities. This is most clearly visible by creating a presence/absence matrix that indicates whether a taxon is present in a sample. Each column in the matrix corresponds to a different sample and each row to a different taxon. When the rows are sorted by taxon prevalence and the columns by richness, as in Figure \ref{fig:nest}b, one can visually verify that the taxa composing the low-diversity communities are also present in most of the higher-diversity communities. 

One possible cause of both these patterns is that microbes require more energy intake to survive in harsher environments\cite{hoehler2013microbial}. For example, powering chaperones to prevent protein denaturation and running ion pumps to maintain pH homeostasis both require significant amounts of ATP. We hypothesized that varying energy demands could explain the patterns observed in the EMP since they would directly alter the severity of environmental filtering.

In the MiCRM, the energetic costs of reproduction are encoded in the model parameter $m_i$, which is the minimal per-capita resource consumption required for net population growth of species $i$ (see Methods for full model equations). The $m_i$ are sampled from a Gaussian distribution with mean 1 and standard deviation 0.01. To  vary the harshness of an environment, we added an environment-specific random number $m_\mathrm{env}$ to the $m_i$ of all species that colonized  a given environment. A large  $m_\mathrm{env}$ corresponds to harsh environments with increased energetic demands whereas small or negative $m_\mathrm{env}$ corresponds to energetically favorable environments.  To mimic the variability in environmental harshness in the EMP, for each community we randomly drew $m_\mathrm{env}$ uniformly between $-0.5$ and $9.5$.

To selectively test the effect of an energy demand gradient on both alpha (within-sample) and beta (between-sample) diversity, we stochastically colonized 300 simulated communities of 150 species each from a regional pool of 180 species with a chemistry of 90 metabolites. We supplied each community with a constant flux of the same resource type.  As discussed above, each of the 300 simulated communities was also assigned a random $m_\mathrm{env}$ to mimic the effects of environmental harshness on growth rates. The results from this simulation are shown in Figure \ref{fig:EMP} and in the bottom left panel of Figure \ref{fig:nest}. The same simulation correctly captures both the richness/harshness correlation and the nestedness of the EMP data, suggesting that these large scale patterns may have a simple origin.

Given the way we have modeled the harshness variations, the link with diversity is not very surprising, because a sufficiently high maintenance cost can make it impossible for a species to survive on a given resource supply, regardless of the surrounding community structure. This pattern is thus guaranteed to occur in any simulation sharing this basic structure. The shape of the richness/harshness relationship does depend on modeling choices, however. We found that diversity loss happens more quickly when the incoming energy is divided among all possible resource types before being supplied to the system, as shown in Figure \ref{fig:EMP}. In this case it can happen that no single species is able to harvest a sufficient number of distinct nutrient sources to meet its maintenance cost, and the whole community goes extinct. In the original simulations, by contrast, the surviving species at high harshness levels satisfy most of their energy requirements by directly consuming the externally supplied resource, with the metabolic byproducts supply sufficient niche differentiation for multiple species to coexist. 

To explore the ecological origins of the nested pattern in more depth, we ran additional simulations in a different regime of community assembly. Instead of modulating diversity with varying levels of selection pressure, we tried varying the degree of dispersal limitation. In the new scenario, each community faced identical environmental conditions, but the initial number of species from the regional species pool allowed to colonize the community was randomly chosen, from 1 to the maximum possible value of 180. In these new simulations, shown in the bottom right panel of Figure \ref{fig:nest}, the nestedness vanishes. The reason for this is that in many environments, only a few species colonize the community resulting in many metabolic niches being unoccupied. We also ran simulations where both $m_\mathrm{env}$ and the initial number of species varied from site to site, and obtained an intermediate degree of nestedness, as shown in Supplementary Figure S1. Collectively, these simulations suggest that nestedness in cross-sectional data may be a sign of selection-dominated community assembly. 

We emphasize that these simulations themselves do not confirm the hypothesis that energy gradients are the driver of the observed patterns. Temperature and pH affect microbes in many other ways that are not included in our minimal model. But our simulations do show that accounting for increased energetic costs associated with harsh environments can reproduce the large-scale patterns observed in the EMP even in the absence of any metabolic or taxonomic structure. Additionally, one ecological factor that seems crucial for reproducing these patterns is dispersal. The nestedness seen in the EMP requires that ecological dynamics are dominated by selection rather than stochastic colonization due to dispersal limitations.

\subsection*{ Metabolic Structure and Species Abundance Curves}

\begin{figure*}[t]
	\centering
	\includegraphics[width=16cm]{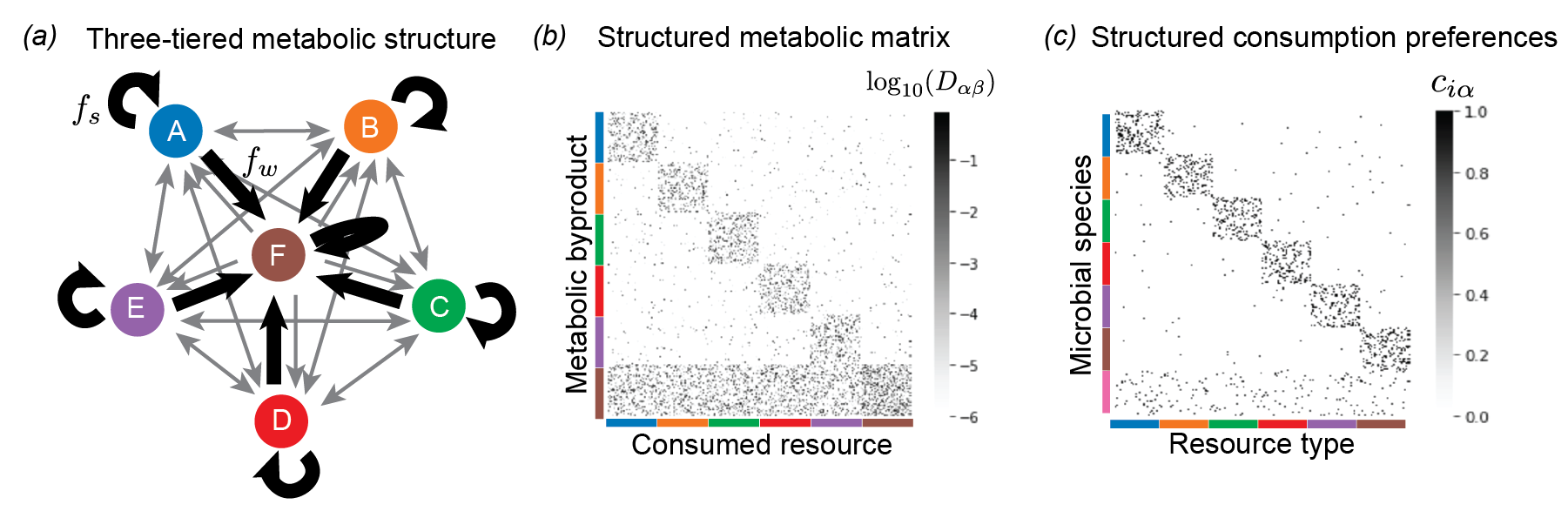}
	\linespread{1}\selectfont{}
	\caption{{\bf Incorporating metabolic and taxonomic structure.} \emph{(a)} Three-tiered secretion model used for simulating human and marine microbiomes. $M=300$ resource types are grouped into $T = 6$ classes of equal size, labeled A through F. These groups represent different kinds of metabolites, e.g. lipids, sugars, amino acids, etc.. Group F is the ``waste'' class, containing common byproducts generated by many metabolic pathways, e.g., carboxylic acids. A fraction $f_s$ of the byproduct flux from metabolism of a given resource is partitioned among resources of the same class. A fraction $f_w$ of the flux is partitioned among ``waste'' resources (class F). The rest of the flux is nonspecifically partitioned among all the other classes. In all simulations shown here, $f_s = f_w = 0.45$. \emph{(b)} Heatmap of a metabolic matrix $D_{\alpha\beta}$ encoding the three-tiered secretion model. \emph{(c)} Taxonomic structure used for human and marine microbiome simulations. Microbial species are grouped into ``families,'' with each family specializing in a different resource class. Specialist families allocate a fraction $q$ of their consumption capacity to their favored resource class. In all the simulations shown here, $q = 0.9$. There is also a generalist family whose preferences are uniformly sampled across all resource types.} 
	\label{fig:struct}
\end{figure*}

In order to reproduce more complex ecological patterns observed such as those observed in the EMP, we  incorporated additional metabolic and taxonomic structure into our model \cite{Goldford2018, Marsland2019a, marsland2019community}, as illustrated in Figure \ref{fig:struct}. The basic idea is to recognize the fact that metabolites often belong to  distinct groups with different metabolic properties ( e.g. lipids, sugars, amino acids, etc.). In most of our simulations, we introduce $T=6$ groups labeled $A-F$ representing these metabolic classes, with  $F$ a  special   ``waste'' class''  which mimics commonly produced metabolic byproducts (i.e. carboyxlic acids for fermentative and respiro-fermentative bacteria). To incorporate this structure in our metabolic matrix we introduce a three-tiered secretion model where: a fraction $f_s$ of the byproduct flux from metabolism of a given resource is partitioned among resources of the same class, a fraction $f_w$ of the flux is secreted as ``waste'' resources (class F), and the rest of the flux is nonspecifically partitioned among all the other classes. 

Different taxonomic families often have distinct resource preferences. For example, it is well known that the bacteria from the taxonomic family  \emph{Enterobacteriaceae} to which \emph{E. coli} belongs preferentially consume sugars.  To reflect such taxonomical preferences in our model, microbial species are grouped into ``families,'' with each family specializing in a different resource class. Specialist families allocate a fraction $q$ of their consumption capacity to their favored resource class. In all the simulations shown here, $q = 0.9$ meaning that specialist families derive $90\%$ of their resources from their preferred resource class. In addition to these specialists, we know that certain microbial families behave as generalists with no strong metabolic preferences across resource types. To model this, we introduce a generalist family whose preferences are uniformly sampled across all resource types.

\begin{figure*}[t]
	\centering
	\includegraphics[width=16cm]{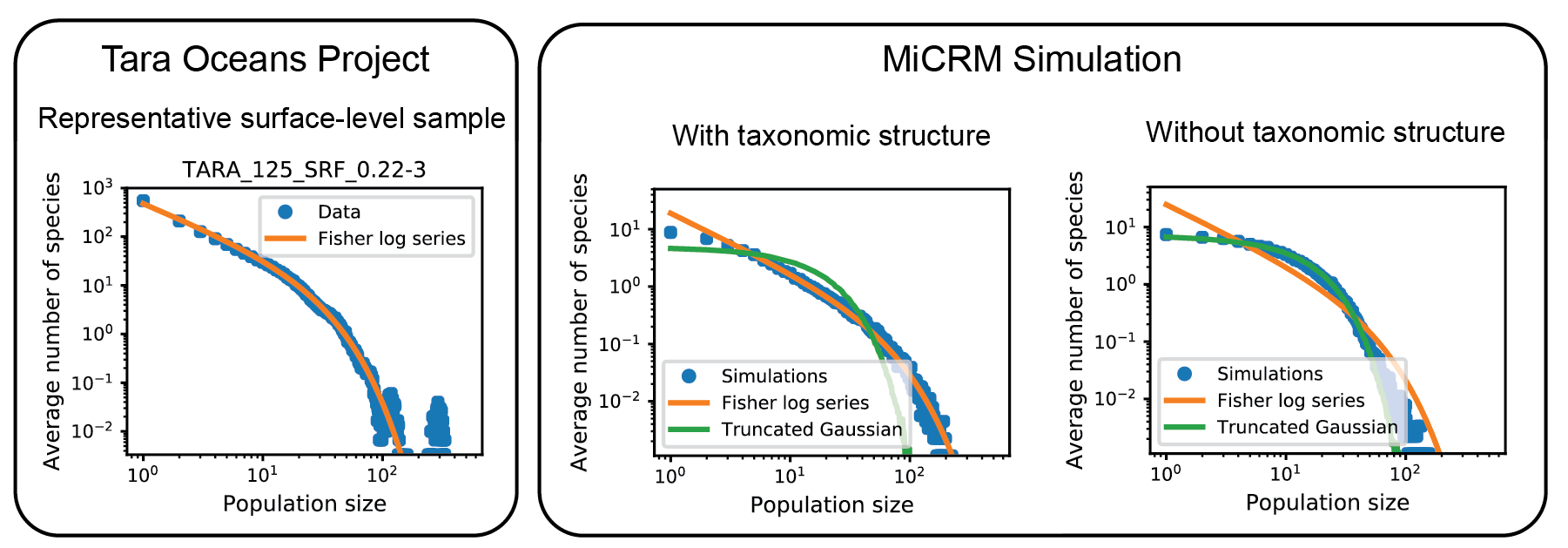}
	\linespread{1}\selectfont{}
	\caption{{\bf Metabolic and taxonomic structure give rise to Fisher log series} Left: Tag sequence count distribution for a representative sea surface sample from the Tara Oceans Project. Data was subsampled 300 times at a depth of 10,000 reads (out of 129,135 in the original sample), and species with 5 reads or less in the raw data were treated as extinct for the purpose of computing the Fisher log series parameters (see Methods). Right: Abundance distributions for simulated communities. 1,000 individuals were sampled from each of 900 simulated communities, with environments and colonization as described for the ``Simple Environments'' panel of Figure \ref{fig:HMP} below. Each point is an average over all 900 communities of the number of species with a given number of individuals. All simulations were performed with the metabolic structure described in Figure \ref{fig:struct} above. The left-hand panel also incorporated taxonomic structure, with different families specializing in different resource classes, with specialization level $q=0.9$. The right-hand panel did not have taxonomic structure ($q = 0$), and consumption preferences for all species were sampled from the same Bernoulli distribution. Green curve (``Truncated Gaussian'') comes from assuming that species' invasion fitness are sampled from a Gaussian distribution, and that population sizes for surviving species are proportional to the invasion fitness, while species with negative invasion fitness go extinct.}
	\label{fig:AD}
\end{figure*}

One commonly employed analysis tool for understanding community structure are species abundance curves. A species abundance curve is made by plotting the number of species present in a sample on the y-axis and the number of individuals or population size on the x-axis. In may ecosystems, it is known that species abundance curves are well fit by a Fisher log series \cite{fisher1943relation, magurran2005species}. Unlike Gaussian distributions or other normal-distribution derived variants such as truncated Gaussians, the Fisher log series has a long tail, reflecting the preponderance of rare species in these ecosystems. As shown in Figure \ref{fig:AD}, the Fisher log series also gives a good fit to species abundance data on ocean microbial communities from the Tara Oceans dataset \cite{Sunagawa2015}. Simulation data generated using the MiCRM with taxonomic structure also result in long-tailed species abundance distributions that are well fit by a Fisher log series. However, these tails disappear in simulations of the MiCRM lacking  metabolic and taxonomic structure. In this case, the species abundance curve were better described using a truncated Gaussian, consistent with theoretical predictions \cite{Advani2018,mehta2018constrained}. These simulations show that the long tailed species abundance curves seen in most ecosystems are compatible with an equilibrium niche model, provided a sufficient level of taxonomic structure, and do not necessarily require neutrality \cite{hubbell_unified_2001} or chaotic dynamics \cite{pearce2019stabilization}.

\begin{figure*}[t]
	\centering
	\includegraphics[width=16cm]{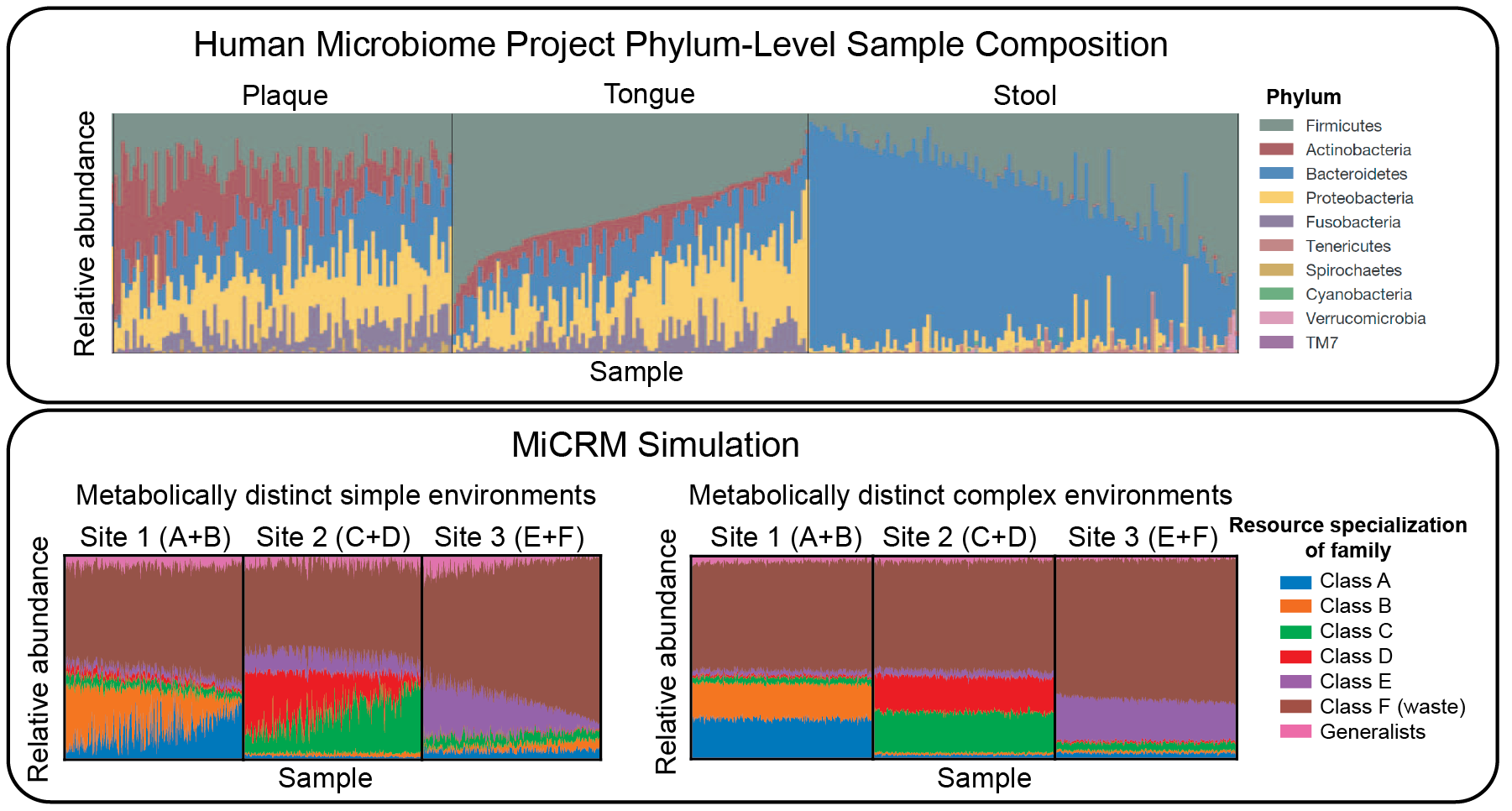}
	\linespread{1}\selectfont{}
	\caption{{\bf Low-dimensional nutrient supply variation reproduces patterns in human microbiome survey data.} Top: Each column represents one sample from the Human Microbiome Project (HMP). Colored segments represent relative abundances of different phyla in each community. Reproduced from Figure 2 of the initial  open-access report on the results of the HMP\cite{HMP}. Bottom: Each column represents one of 900 simulated samples, each stochastically colonized with 2,500 species from a regional pool of 5,000 species, comprising seven metabolically distinct families. Colored segments represent relative abundances of the seven families defined in Figure \ref{fig:struct}. Each of the three ``body sites'' was supplied with resources from a different pair of resource classes, with total nutrient supply fixed. In the first set of simulations (left), one resource from each class was supplied, and the ratio of the two supply rates was randomly varied from sample to sample. In the second set (right), all resources from each class were supplied, with randomly chosen supply rates for each sample, normalized to keep the total supply fixed. The brown family present in all three environments specializes in the typical byproducts (e.g., carboxylic acids) generated from all the other resource classes. Within each body site, samples are sorted by relative abundance of this family. See main text and Methods for simulation details.}
	\label{fig:HMP}
\end{figure*}

\begin{figure*}[t]
	\centering
	\includegraphics[width=16cm]{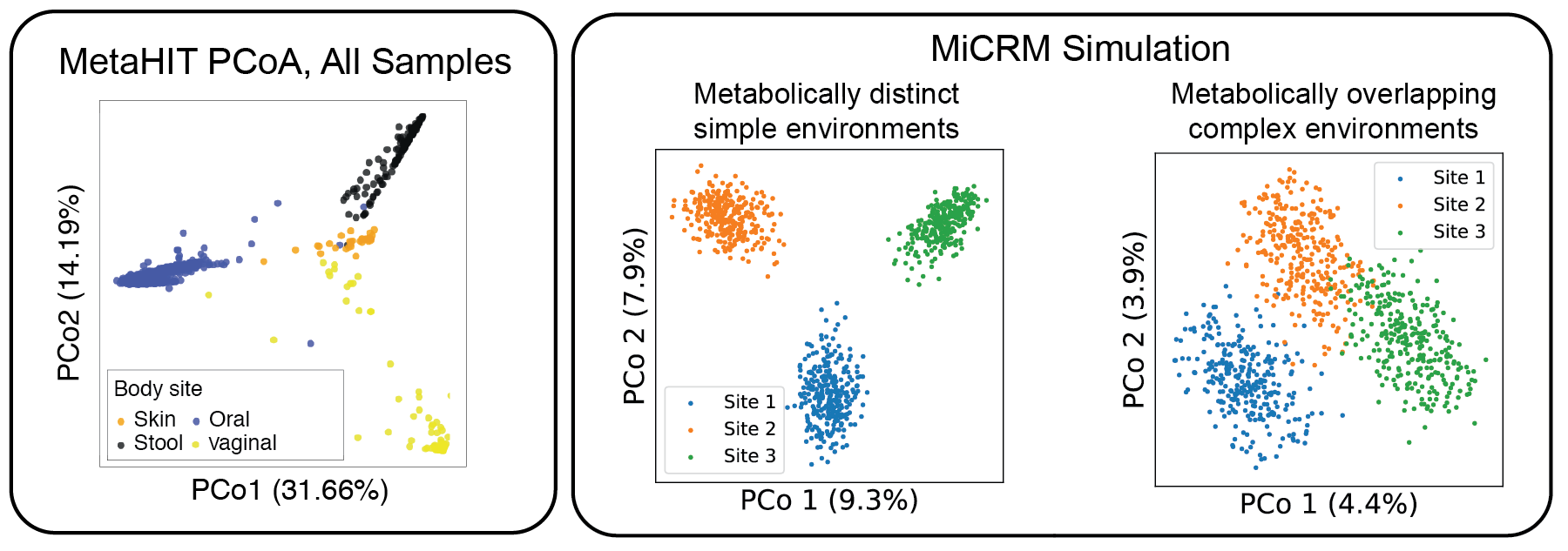}
	\linespread{1}\selectfont{}
	\caption{{\bf Correlations between inter-site nutrient variation and metabolic structure affect distinguishability of body sites.} Left: Principal coordinate analysis (PCoA) of MetaHIT OTU-level community compositions, using the Jensen-Shannon distance metric. Data points are colored by the body site from which the sample was taken. Reproduced with permission from Figure 1 of \cite{costea2018enterotypes}. Right: Jensen-Shannon PCoA of species-level compositions of the simulated communities. In the first set of simulations (left), the nutrients supplied to different body sites come from different resource classes. This is the same set of simulations used for the left-hand panel of Figure \ref{fig:HMP}, but similar results are obtained if the simulations of the other panel are used instead, or if consumption preferences are uniformly random with no taxonomic structure (See Supplementary Figure S2). In the second set of simulations (right), each environment is supplied with a randomly chosen set of resource types, with each site being supplied with about one third of the 300 possible resources. See main text and Methods for simulation details.}
	\label{fig:PCoA-all}
\end{figure*}

\begin{figure*}[t]
	\centering
	\includegraphics[width=16cm]{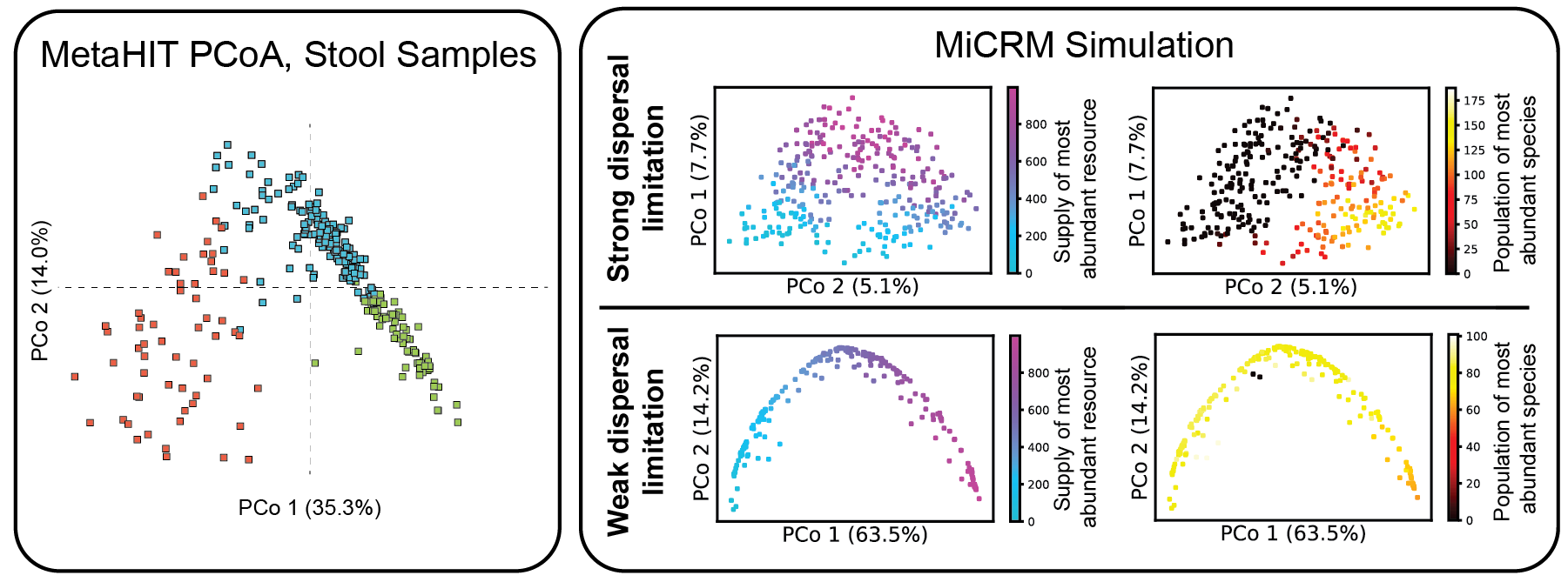}
	\linespread{1}\selectfont{}
	\caption{{\bf Pattern in ordination of compositions from single body site admits of multiple explanations.} Left: Jensen-Shannon PCoA of MetaHIT stool samples, showing a characteristic `U' shape that has been observed in many independent studies. Colors indicate three hypothesized enterotypes, which we do not discuss here.  Reproduced with permission from Figure 3 of \cite{costea2018enterotypes}. Right: Jensen-Shannon PCoA of simulated samples from Body Site 1 under two different levels of dispersal limitation. In the first (top), each community was initialized with 2,500 randomly chosen species out of the regional pool of 5,000. The communities display a continuous gradient in the population size of the most abundant species (over all samples) along the `u' shape from one end to the other. In the second (bottom), each community started with 4,900 species. These communities display a continuous gradient of environmental conditions along the `U' shape from one end to another.}
	\label{fig:PCoA-single}
\end{figure*}

\subsection*{Patterns in the HMP can be explained by environmental filtering and competition}
The Human Microbiome Project is a large-scale survey of the microbial communities that reside in and on the human body \cite{HMP}. The  HMP was supplemented by the smaller MetaHIT project which focused on sequencing  fecal metagenomes from multiple individuals. Initial analysis of the HMP and MetaHIT results on the human microbiome revealed three major patterns, displayed in the top half of Figures \ref{fig:HMP},  \ref{fig:PCoA-all}, and \ref{fig:PCoA-single}. First, for a given body site different individuals had very different community compositions (see Fig. \ref{fig:HMP}) . Even at the phylum level, the relative abundances of dominant taxa varied dramatically from sample to sample \cite{HMP}. But samples from different body sites still typically differed more than samples from the same body site, leading to the second pattern, shown in Figure \ref{fig:PCoA-all} of clustering of microbial communities by body site across individuals \cite{HMP,qin2010human}. Finally, the gradients in the relative abundances of the dominant taxa in a given body site across individuals were also visible in dimensional reductions of more fine-grained (genus-level) community composition, producing the third pattern shown in Figure \ref{fig:PCoA-single}. 

One factor associated with the compositional gradient is the host's typical diet \cite{costea2018enterotypes,gorvitovskaia2016interpreting}. Different kinds of externally supplied nutrients, such as fibers and proteins, are thought to encourage growth of different microbial taxa. For this reason, we hypothesized that the patterns in the HMP may arise from heterogeneity in the resources available in different environments.  It is clear that reproducing such patterns requires assuming some minimal level of taxonomic and metabolic structure. For this reason, in our simulations we divided resources into six resource classes and species into six families, with each family specializing in one resource class, as illustrated in Figure \ref{fig:struct} and described above.

We first constructed \emph{metabolically structured simple environments} where there were only two externally supplied resources. In particular, each of the three ``body sites'' was supplied with a unique  pair of resources from distinct resource classes (i.e. body site 1 was supplied with a resource from class A and a resource from class B, body site 2 with a resource from class C and a resource from class D, and body site 3 with a resource from class E and a resource from class F). We modeled variability in the availability of resources across individuals at a fixed body site by changing the ratio of the two supplied resources while holding the total supplied energy fixed (see Methods). We also created \emph{metabolically structured complex environments} where each body site was supplied with 50 external resources from each of the two resource classes while holding the total supplied energy fixed (i.e. body site was supplied with all 50 resources from class A and all 50 resources from class B, body site 2 with all 50 resources from class C and all 50 resources from class D, and body site 3 with all 50 resources from class E and all 50 resources from class F).

	To mimic the scale of the actual microbiome data, we generated a regional pool of 5,000 species (approximately the number of OTU's identified in the HMP \cite{HMP}), and stochastically colonized 300 samples per body site with 2,500 species each. Figure \ref{fig:HMP} shows the resulting patterns for simple and complex environments. For simple environments, our simulations mimic the broad range of compositions found in the data including gradients in the dominant families present at each of the body sites. In contrast, for complex environments we see that the relative abundance of different families stays almost constant across individuals for each body site. This suggests that the patterns found in Fig \ref{fig:HMP} may reflect the combined effects of environmental filtering and competition between species in the presence of a few dominant externally supplied resources.

We used the data from simulations on metabolically structured simple environments to perform a PCoA across body sites as in the MetaHIT data. As can be seen in Figure \ref{fig:PCoA-all}, these simulations recapitulated the pattern seen in real microbial communities. We found that this clustering by body site depended strongly on the fact that different body sites had metabolically distinct resources. For example, if we instead considered \emph{metabolically unstructured complex environments} where each body-site was supplied with 100 randomly chosen distinct resources regardless of resource class, the clusters were no longer fully separable on a two-dimensional PCoA (right most graph in Figure \ref{fig:PCoA-all}). This suggests that the clustering of human microbiomes according to body-sites likely reflects the fact that these body sites have metabolically distinct environments that result in different patterns of byproduct secretion.

We also investigated the ability of our model to reproduce the U-shaped curves observed in PCoA of communities at a single body site (see Figure \ref{fig:PCoA-single}). We found that we could reproduce this pattern using the same simulations used in Figure \ref{fig:PCoA-all} to understand metabolically structured simple environments. With the level of dispersal limitation used in these simulations, the U shape primarily results from the stochastic presence or absence of the most abundant species. If we reduce dispersal limitation, however, by initializing each community with nearly all of the possible species, the U shape directly reflects the low-dimensional variability of resource supply in the simple environments. It remains unclear which, if any, of these two explanations corresponds to the pattern in the gut microbiome, whose significance is a matter of ongoing controversy \cite{Arumugam2011,costea2018enterotypes,gorvitovskaia2016interpreting}.

\subsection*{Dissimilarity-overlap patterns reflect shared environments not universal dynamics}

\begin{figure*}[t]
	\centering
	\includegraphics[width=16cm]{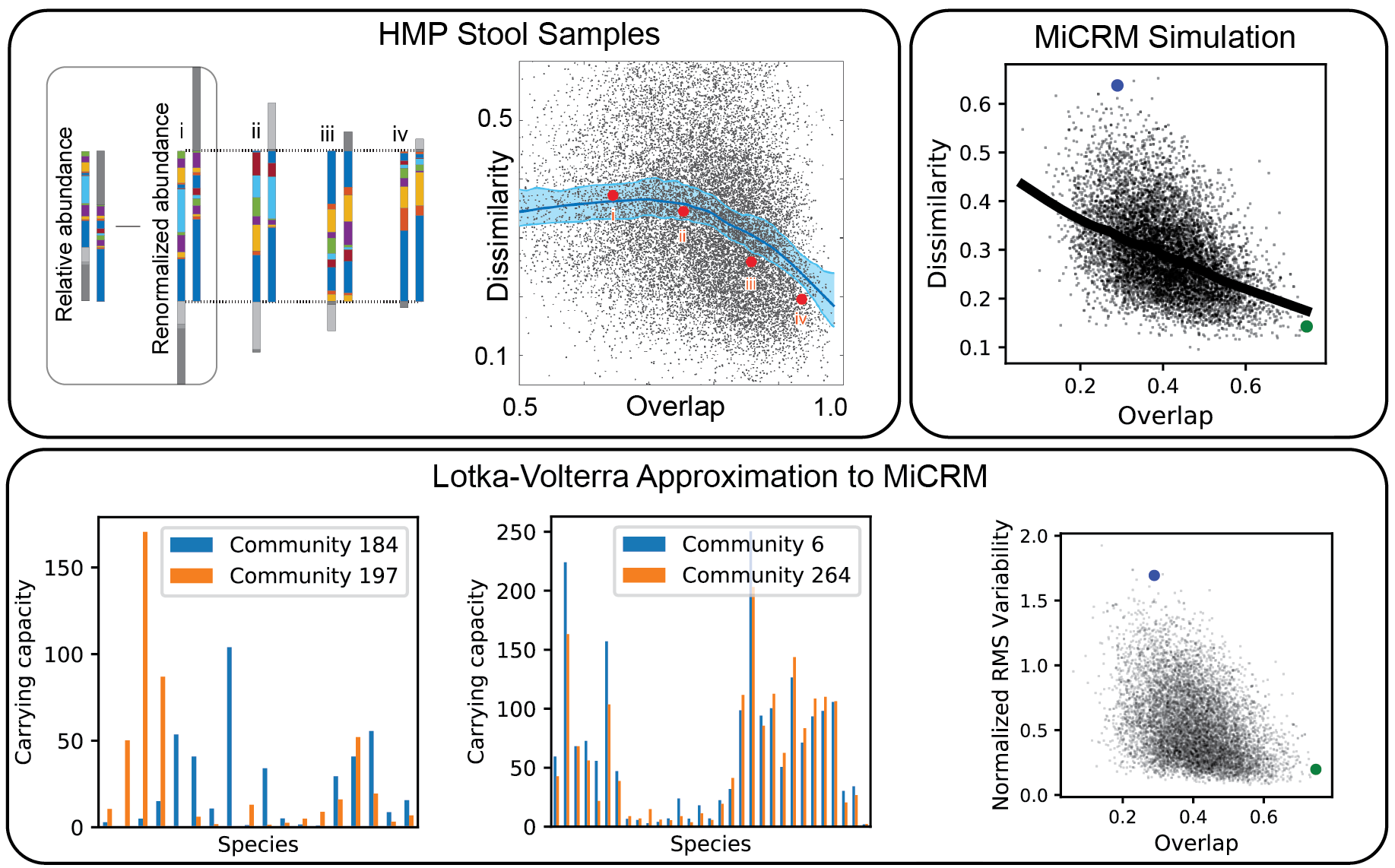}
	\linespread{1}\selectfont{}
	\caption{{\bf Host-specific dynamics are compatible with dissimilarity-overlap correlation.} Top left: The composition of pairs of samples can be compared in two independent ways: ``overlap'' measures the fraction of each sample comprised by species common to both, and ``dissimilarity'' measures how different the relative abundance profiles are within this shared pool. The four pairs shown here have increasing overlap and decreasing dissimilarity from left to right, corresponding to the four points indicated in the scatter plot. Dissimilarity and overlap are plotted for 17,955 pairs of stool samples from the HMP, analyzed at the genus level. Solid line is a Lowess smoothing of the data, and red points correspond to the sample pairs illustrated in the first panel. Reproduced with permission from \cite{bashan2016universality}. Top right: Dissimilarity and overlap for 10,000 pairs of simulated samples from the metabolically distinct simple environments of fig.~\ref{fig:HMP}, with one resource supplied from class A and one from class B. Solid line is a Lowess smoothing of the data. Blue and green points correspond to two representative pairs of communities selected for further analysis in the bottom panel. Bottom: For each sample, the population dynamics near the steady state was approximated with a generalized Lotka-Volterra model. Effective carrying capacities and interaction coefficients computed from the mechanistic model parameters together with the population sizes and resource abundances, as described in the Methods. We have plotted the carrying capacity of each species for two representative pairs of communities with low (left) and high (center) overlap. These pairs are indicated in the scatter plots by blue and green points, respectively. We also show the normalized root-mean-square variability in carrying capacity for all 10,000 sample pairs (right).}
	\label{fig:DOC}
\end{figure*}

Another pattern obtained from a more recent analysis of the HMP data is an anti-correlation between overlap and dissimilarity of pairs of communities from a given body site (see Figure \ref{fig:DOC} and \cite{bashan2016universality} for details). Due to both stochastic colonization and variable environments, there are usually many species in one sample that are not present in the other. Different pairs of samples overlap to different degrees, and this overlap can be measured in terms of the ratio of the combined population of the shared species to the total population of the two samples. If one focuses on the subset of species that are shared, one can also compare the relative abundance distributions of the two samples within this shared pool, as illustrated in Figure \ref{fig:DOC}, using standard measures of dissimilarity such as the Jensen-Shannon divergence (see \cite{bashan2016universality} for detailed discussion). These two quantities are not intrinsically related, as can be seen by evaluating them over a randomly generated table of abundances \cite{bashan2016universality}. 

This analysis was initially proposed as a way of distinguishing between ``universal'' and ``host-specific'' microbial dynamics. It was argued that if the dynamics of human associated microbial communities were universal, different individuals could be modeled with the same dynamic parameters and this would be reflected by a negative correlations between dissimilarity and overlap across cross-sectional samples. In contrast, for host-specific dynamics each individual would have their own kinetic parameters and the dissimilarity and overlap would be uncorrelated. This interpretation has been disputed, however, with numerical simulations of Lotka-Volterra type models showing that a negative correlation can result from host-specific dynamics in the presence of stochasticity, sampling errors, or environmental gradients \cite{kalyuzhny2017dissimilarity}.

We re-ran the analysis of  \cite{bashan2016universality} on our simulated HMP data discussed above and found that the dissimilarity and overlap were negatively correlated at a single body site just as in the real gut microbiome data (see Figure \ref{fig:DOC}, top right). However, this correlation was absent if we analyzed pairs of samples from distinct body sites, indicating that this signature likely arises due to the fact the all the communities at a given body site exist in a similar external environment (see Supplementary Figure S3). Importantly, our simulations show that the negative dissimilarity-overlap correlation observed \cite{bashan2016universality} can be found even in the absence of universal dynamics since environments with different amounts of externally supplied resources generically give rise to communities with different 
ecological dynamics. Instead, our simulations suggest that the negative correlation between overlap and dissimilarity found in the HMP may reflect the fact 
that communities at a given body-site experience similar but not identical environments.

As a further check,  we approximated the population dynamics near the steady state using a generalized Lotka-Volterra model (see Methods). This allowed
us to explicitly calculate the effective carrying capacities and interaction coefficients for each community. The bottom row of Figure \ref{fig:DOC} shows the carrying capacity for two pairs of communities: one pair where the two communities in the pair have a  high overlap and another where the communities in the pair have a low overlap. The carrying capacities of species in the high-overlap communities are extremely similar where as the carrying capacities of the low overlap communities are very different from each other. This provides strong evidence for the important role played by environmental filtering in producing the dissimilarity-overlap pattern observed in the HMP data.

\subsection*{``Modular assembly'' of microbial communities}

\begin{figure*}[t]
	\centering
	\includegraphics[width=16cm]{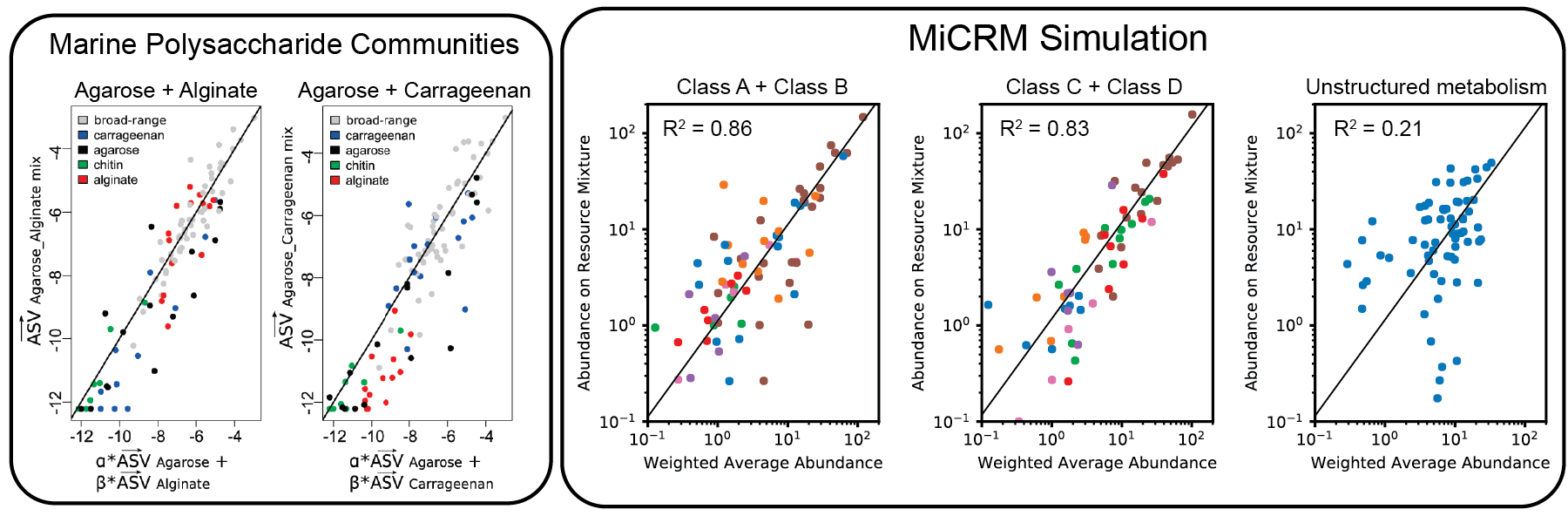}
	\linespread{1}\selectfont{}
	\caption{{\bf Modularity of community assembly.} Left: In the experiments reported in \cite{enke2018modular}, synthetic beads composed of different kinds of polysaccharides, including agarose, alginate and carrageenan, were incubated with coastal seawater and colonized by the marine bacteria resident in the seawater sample. 16S rRNA amplicon profiling was performed for communities grown on beads composed of a single kind of polysaccharide, as well as mixtures of two kinds of polysaccharides. Relative abundances of amplicon sequences variants for two different mixtures (Agarose/Alginate and Agaraose/Carrageenan) are plotted versus a weighted average of the relative abundances on the pure beads. Solid lines are fits to a linear mixture model, with R$^2$ of 0.84 and 0.74, respectively. Right: Abundance of each species in simulated communities supplied with mixtures of two resource types, plotted against the average of the abundances for communities supplied with just one of the resource types, with the total energy supply held constant. For the first two panels, all other parameters are the same as for the human microbiome simulations of Figure \ref{fig:HMP}, except that each sample is initialized with all 5,000 species from the regional pool. Titles indicate the class labels of the two supplied resources for each scenario, and species are colored by metabolic family following Figure \ref{fig:struct}. In the third panel, simulations were run with the same number of resources and species, but with all resources assigned to the same resource class, eliminating all metabolic and taxonomic structure. Solid lines are predictions of the additive model where the abundance in the mixture equals the average of the abundances in the single-resource condition. The R$^2$ score of this model is also shown in each panel.}
	\label{fig:modular}
\end{figure*} 

Our analysis of our synthetic HMP data also shows a new pattern: the family-level composition of each community along the nutrient gradient is approximately a linear combination of the compositions of the two extreme communities on the gradient (Figure \ref{fig:HMP}d). Quantitatively, if $N_i^1$ are the population sizes for each family $i$ in the community supplied with a flux $\kappa_1$ of resource type 1 alone, and $N_i^2$ are the population sizes in the community supplied with flux $\kappa_2$ of resource type 2 alone, then the community supplied with flux $(1-\alpha)\kappa_1$ of resource type 1 and flux $\alpha$ of resource type 2 has population sizes approximately equal to $(1-\alpha) N_i^1 + \alpha N_i^2$. This ``additivity'' of communities from different environments can also be seen at the species level, by plotting the actual population sizes $N_i^{\rm mix}$ versus the weighted average prediction, as shown in Figure \ref{fig:modular}.

Since this analysis is performed at the species level, meaningful taxonomic groupings are no longer necessary, so we also checked for additivity in simulations where communities lack any metabolic or taxonomic structure (i.e with unstructured metabolic and consumer preference matrices). Figure \ref{fig:modular} shows that the population sizes in the mixed-resource communities are not well predicted by the weighted average of single-resource communities, with an $R^2$ of 0.21. This suggests that metabolic and taxonomic structure are necessary to see this additive pattern.

This pattern is difficult to test for in field survey data, since there are many additional factors besides diet that vary from sample to sample, many of which may themselves be correlated with diet. Studying this kind of effect requires controlled experiments where the variable of interest can be systematically varied. One recent experiment colonized seawater communities on small beads composed of different kinds of carbohydrates, which served as the sole externally provided carbon source for the community \cite{enke2018modular}. This simplified scenario reflects the conditions of our minimal model more closely, where only one or two nutrient types were externally supplied. The authors of \cite{enke2018modular} compared the weighted average of population sizes from communities grown on two different carbon sources, such as agarose and alginate, and the corresponding population sizes from other communities grown on a mixture of the two. They found the same ``additivity'' effect we observe in our simulations, with even stronger $R^2$ values of 0.84 and 0.74 for two different carbon source pairs. They termed this property ``modular assembly'' of microbial communities. Our simulations show that modular assembly may be a generic property of complex microbial communities grown in the presence of multiple metabolically distinct resources.

\section*{Discussion}
We have shown that the Microbial Consumer Resource Model introduced in \cite{Goldford2018} to describe laboratory experiments in synthetic minimal environments can also reproduce a wide range of experimentally observed patterns in survey data such as the HMP and EMP including harshness/richness correlations, nestedness of community composition, compositional gradients, and dissimilarity/overlap correlations. The MiCRM provides a systematic way of exploring the effect of stochastic colonization, resource competition, and metabolic crossfeeding on large-scale observables. By randomly sampling parameters from well-defined probability distributions, we combine a sufficient level of mechanistic detail to make the parameters physically meaningful, while keeping the number of parameters small enough for systematic investigation of the factors that control different patterns.  

Our numerical results complement recent theoretical works suggesting that complex ecosystems may still be well described by random ecoystems \cite{barbier2018generic, cui2019diverse}, suggesting the essential ecology of diverse ecosystems may be amenable to analysis using techniques from statistical mechanics and random matrix theory. For these reasons, the MiCRM is well-suited to serve as a minimal model for understanding microbial ecology.

Our analysis also suggests several hypotheses relating mechanism to large scale patterns in both the EMP and HMP. We have shown that it is possible to reproduce the richness/harshness correlation and the nestedness of the EMP data by assuming that harsh environments pose an additional energetic cost to organisms. This is true even when communities are grown in otherwise identical environments and lack any taxonomic and metabolic structure. This complements earlier work showing that energy availability is a key driver of community function and structure \cite{marsland2018available}. Our simulations on the HMP suggest that environmental gradients and resource availability result in significant environmental filtering and naturally explain the clustering of microbial communities by body site. 

We have also identified ecological parameters that can break the observed patterns, allowing us to generate hypotheses about the underlying ecological processes: (a) breaking the nestedness pattern (Fig. \ref{fig:nest}) with dispersal limitation allowed us to connect nestedness to a selection-dominated regime in the EMP; (b) the loss of compositional gradients in complex environments (Fig. \ref{fig:HMP}) led us to hypothesize that a small number of dominant resource types may drive inter-subject variability in the HMP; (c) the degradation of compositional clustering by body site (Fig. \ref{fig:PCoA-all}) in metabolically unstructured environments highlighted the importance of metabolically relevant differences between resource environments in the HMP; (d) breaking the additivity of communities grown on mixed resource supplies (Fig. \ref{fig:modular}) allowed us to connect this pattern to taxonomic and metabolic structure in the microbial species.

Great care has to be taken when interpreting large-scale patterns. For example, the negative correlation between dissimilarity and overlap observed in HMP data in \cite{bashan2016universality} may be indicative of the fact that body-sites across individuals have similar environments rather than a much stronger claim of universal dynamics in the human microbiome. Our work also suggests that many large scale patterns may occur generically across different environmental settings. For example, we have shown that the additivity observed in our synthetic HMP data is also observed in ocean communities grown on synthetic carbon beads \cite{enke2018modular}, suggesting modular assembly may be a generic property of communities grown in environments with metabolically distinct resources.

The analysis presented here shows that it is possible to qualitatively reproduce patterns seen in large-scale surveys such as the EMP and HMP
using a simple minimal model. An interesting area of future research is to move beyond qualitative comparisons and ask how minimal models and large scale simulations can be quantitatively compared to large-scale genomic surveys. This problem is especially challenging given the large number of parameters, environments, and experimental designs that must be explained. One potential avenue for doing this is to use statistical methods such as Approximate Bayesian Computation (ABC) \cite{csillery2010approximate}. In ABC, the need to exactly calculate complicated likelihood functions is replaced with the calculation of summary statistics and numerical simulations. In this way, it may be possible to quantitatively relate mechanistic details at the level of microbes to community level patterns observed in large-scale surveys.

\section*{Methods}
All synthetic data was generated using the Microbial Consumer Resource Model previously described\cite{marsland2018available, marsland2019community}. 
We found the fixed points of the dynamics for each community using the Python package Community Simulator \cite{marsland2019community}:\\ \url{https://github.com/Emergent-Behaviors-in-Biology/community-simulator}.\\ Principal Coordinate Analysis was performed on the simulated HMP data using the Python package scikit-bio \url{http://scikit-bio.org/}. The pairwise distance matrix was generated using standard scipy commands with the Jensen-Shannon distance metric. Dissimilarity-overlap analysis was performed on the simulation data following the procedure described in \cite{bashan2016universality}. All simulation data and analysis scripts are available at \url{https://github.com/Emergent-Behaviors-in-Biology/microbiome-patterns}.

\subsection*{MiCRM Dynamical Equations}

\begin{table}
	\begin{tabular}{|c|l|}
		\hline
		$N_i$ & population density of species $i$ (individuals/volume)\\
		\hline
		$R_\alpha$ & Concentration of resource $\alpha$ (mass/volume)\\
		\hline
		$c_{i\alpha}$ & Uptake rate per unit concentration of resource $\alpha$ by species $i$ (volume/time)\\
		\hline
		$D_{\alpha\beta}$ & Fraction of byproducts from resource $\beta$ converted to $\alpha$ (unitless)\\
		\hline
		$g_i$ & Conversion factor from energy uptake to growth rate (1/energy)\\
		\hline
		$w_\alpha$ & Energy content of resource $\alpha$ (energy/mass)\\
		\hline
		$l_\alpha$ & Leakage fraction for resource $\alpha$ (unitless)\\
		\hline
		$m_i$ & Minimal energy uptake for maintenance of species $i$ (energy/time)\\
		\hline
		$\kappa_\alpha$ & External supply of resource $\alpha$ (mass/volume/time)\\
		\hline
		$\tau_R$ & Timescale for resource dilution (time)\\
		\hline
	\end{tabular}
\caption{\bf Definitions and units for mechanistic parameters.}
\label{tab:param}
\end{table}

We consider the dynamics of the population densities $N_i$ of $S$ microbial species and the abundances $R_\alpha$ of $M$ resource types in a well-mixed system, governed by the following set of ordinary differential equations:
\begin{align}
\frac{dN_i}{dt} &= g_i N_i \left[ \sum_{\alpha}w_\alpha (1-l_\alpha) c_{i\alpha} R_\alpha - m_i \right]\\
\frac{dR_\alpha}{dt} &= \kappa_\alpha - \tau_R^{-1} R_\alpha - \sum_{i} c_{i\alpha} N_i R_\alpha \nonumber\\
&+ \sum_{i,\beta} D_{\alpha\beta} \frac{w_\beta}{w_\alpha} l_\beta c_{i\beta}N_i R_\beta.
\label{eq:MiCRM}
\end{align}
For this study, the conversion factors $g_i$ from energy uptake to population growth were all set to 1, as were the resource qualities $w_\alpha$ and the resource dilution rate $\tau_R^{-1}$. The leakage fractions $l_\alpha$ govern how much of each consumed resource is released into the environment as metabolic byproducts, and was set to 0.8 for all $\alpha$. See Table \ref{tab:param} for list of all parameters and units.

\subsection*{Random sampling of consumer preference matrix and metabolic matrix}
As noted in the Introduction, modeling highly diverse communities such as microbiomes requires a large number of free parameters. For example, the simulations with 5,000 species performed here required choosing over a million parameter values. In order to explore the typical phenomena produced by our model, we sampled the parameters randomly. Under the sampling scheme described in this section, the model is fully defined by a choice of just twelve parameters, listed in Table \ref{tab:meta}. 

\begin{table}
	\begin{tabular}{|c|l|}
		\hline
		$M$ & Number of resources\\
		\hline
		$T$ & Number of resource classes\\
		\hline
		$S_{\rm tot}$ & Number of microbial species in regional pool\\
		\hline
		$F$ & Number of specialist families\\
		\hline
		$S$ & Number of microbial species initially present in each local community\\
		\hline
		$\mu_c$ & Mean sum over a row of the preference matrix $c_{i\alpha}$\\
		\hline
		$c_0$ & Low consumption level for binary $c_{i\alpha}$\\
		\hline
		$c_1$ & High consumption level for binary $c_{i\alpha}$\\
		\hline
		$q$ & Fraction of consumption capacity allocated to preferred resource class\\
		\hline
		$s$ & Sparsity of metabolic matrix\\
		\hline
		$f_w$ & Fraction of secreted byproducts allocated to ``waste'' resource class\\
		\hline
		$f_s$ & Fraction of secreted byproducts allocated to same resource class\\
		\hline
	\end{tabular}
\caption{\bf Definitions of global parameters used for constructing random ecosystems}
\label{tab:meta}
\end{table}

We choose consumer preferences $c_{i\alpha}$ as follows. We assume that each specialist family has a preference for one resource class $A$ (where $A=1 \ldots  F$) with $0 \le F \le T$, and we denote the consumer coefficients for this family by $c_{i \alpha}^A$. We also consider generalists that have no preferences, with consumer coefficients $c_{i \alpha}^{\mathrm{gen}}$. The $c_{i \alpha}^A$ can be drawn from one of three probability distributions : (i) a Normal/Gaussian distribution, (ii) a Gamma distribution (which ensure positivity of the coefficients), and (iii)  a Bernoulli distribution with binary preference levels. 

The key parameters for constructing all three distributions are the mean $\mu_c$ and the variance $\sigma_c^2$ of the sum $\sum_\alpha c_{i\alpha}$ over a row of the matrix. 

In the current study, we focus on binary preference levels (option iii). In this model, there are two possible values for each $c_{i\alpha}$: a low level $\frac{c_0}{M}$ and a high level $\frac{c_0}{M} + c_1$. For a given choice of $\mu_c$, the parameters $c_0$ and $c_1$ together determine the variance $\sigma_c^2$. The elements of $c_{i \alpha}^A$ are given by
\begin{align}
c_{i \alpha}^A = \frac{c_0}{M} + c_1 X_{i \alpha},
\end{align}
where $X_{i \alpha}$ is a binary random variable that equals 1 with probability
\begin{align}
p_{i \alpha}^A =
\begin{cases}
{\mu_c \over M c_1}\left[1+\frac{M-M_A}{M_A}q\right], & \text{if}\ \alpha \in A \\
{\mu_c \over M c_1}(1-q), & \text{otherwise}
\end{cases}
\label{eq:p}
\end{align}
for the specialist families, and
\begin{align}
p_{i \alpha}^\mathrm{gen}=\frac{\mu_c}{Mc_1}
\end{align}
for the generalists. 

We choose the metabolic matrix $D_{\alpha\beta}$ according to the three-tiered secretion model depicted in Figure \ref{fig:struct} The first tier is a preferred class of `waste' products, such as carboyxlic acids for fermentative and respiro-fermentative bacteria, with $M_w$ members. The second tier contains byproducts of the same class as the input resource (when the input resource is not in the preferred byproduct class). For example, this could be attributed to the partial oxidation
of sugars into sugar alcohols, or the antiporter behavior of various amino acid
transporters. The third tier includes everything else. We encode this structure in $D_{\alpha\beta}$ by 
sampling each column $\beta$ of the matrix from a Dirichlet distribution with concentration parameters
$d_{\alpha\beta}$ that depend on the byproduct tier, so that on average a fraction $f_w$ of the secreted flux goes to the first tier, while a fraction $f_s$ goes to the second tier, and the rest goes to the third. The Dirichlet distribution has the property that each sampled vector sums to 1, making it a natural way of randomly allocating a fixed total quantity (such as the total secretion flux from a given input). To write the expressions for these parameters explicitly, we let $A(\alpha)$ represent the class containing resource $\alpha$, and let $w$ represent the `waste' class. We also introduce a parameter $s$ that controls the sparsity of the reaction network, ranging from a dense network with all-to-all connection when $s \to 0$, to maximal sparsity with each input resource having just one randomly chosen output resource as $s \to 1$. With this notation, we have
\begin{align}
D_{\alpha\beta} &= {\rm Dir}(d_{1\beta},d_{2\beta},d_{3\beta},\dots,d_{M\beta})_\alpha\\
d_{\alpha\beta} &=
\begin{cases}
\frac{f_w}{sM_w}, & \text{if}\, A(\beta) \neq w \text{ and } A(\alpha) = w\\
\frac{f_s}{sM_{A(\beta)}}, & \text{if}\,  A(\beta) \neq w \text{ and } A(\alpha) = A(\beta) \\
\frac{1-f_s-f_w}{s(M-M_{A(\beta)}-M_w)}, & \text{if}\,A(\beta),A(\alpha) \neq w \text{ and } A(\alpha) \neq A(\beta) \\
\frac{f_w+f_s}{sM_w}, & \text{if}\, A(\beta) = w \text{ and } A(\alpha) = w\\
\frac{1-f_w-f_s}{s(M-M_w)}, & \text{if}\, A(\beta) = w \text{ and } A(\alpha) \neq w.
\end{cases}
\end{align}
The final two lines handle the case when the `waste' type is being consumed. For these columns, the first and second tiers are identical. 

\subsection{Solving for uninvadable equilibrium}
We computed the uninvadable equilibrium state of Equations (\ref{eq:MiCRM}) using a novel algorithm inspired by expectation-maximization methods in machine learning. The algorithm is described in detail in \cite{marsland2019community}, and implemented computationally in the Community Simulator package. 

The raw results of the computation have nonzero abundances for all species, due to technical limits on numerical precision in the solver. In all simulations, the abundance distribution was clearly bimodal, with well-separated peaks on a log scale for the surviving vs. extinct species. For purposes of determining species richness, we set the abundance of all species in the ``extinct'' group to zero. Histograms of the raw results are plotted in the accompanying Jupyter notebook, where the choice of threshold for removing extinct species can be directly verified.

The large simulations with 300 resources and 5,000 species pushed the limits of our implementation of the algorithm, and occasionally failed to converge. Before performing further analysis, we directly verified that a true solution had been found by calculating the per-capita growth rate $d\ln N_i/dt$ for all surviving species. A histogram of the maximum value of $|d\ln N_i/dt|$ for each community (on a log scale) shows that most simulations are around $10^{-7}$, with the upper tail reaching to around $10^{-5}$. In the least stable simulation, with $S = 4,900$ and two externally supplied resources, the failed runs form a second cluster around $|d\ln N_i/dt| = 10^{-3}$. To eliminate such runs, we set a threshold for all simulations discarding samples with $|d\ln N_i/dt| \geq 10^{-5}$. For the least stable scenario, 29 of the 900 samples exceeded the threshold, and all others had between 0 and 11.

\subsection{Synthetic data for global biodiversity patterns}
Synthetic data for Figures \ref{fig:EMP} and \ref{fig:nest} was generated using a regional species pool of size $S_{\rm tot}=180$ and $M=90$ potential resources. The elements of the $180\times 90$ consumer preference matrix $c_{i\alpha}$ were sampled from a binary distribution as described above, with $c_0 = 0$, $c_1 = 1$ and $\mu_c = 10$, using only one resource class ($T = 1$) and one consumer family ($F=1$). The $90\times 90$ metabolic matrix $D_{\alpha\beta}$ was sampled from a Dirichlet distribution as described above, with $s = 0.05$. The $m_i$ were sampled from a Gaussian distribution with mean 1 and standard deviation 0.01. A random number from a uniform distribution between -0.5 and 9.5 was added to all the $m_i$'s from each sample. 

The rest of the parameters differed among the three scenarios we simulated, and were chosen as follows:
\begin{itemize}
	\item {\bf Simple environment} (same as ``selection-limited'' in Figure \ref{fig:nest}): Each sample was stochastically colonized with $S = 150$ out of the 180 possible species, and supplied with a single external resource, with $\kappa_1 = 200$ and $\kappa_\alpha = 0$ for all $\alpha \neq 1$.
	\item {\bf Complex environment}: Each sample was stochastically colonized with $S = 150$ out of the 180 possible species, and supplied with all external resources, with $\kappa_\alpha = 200/M = 2.2$ for all $\alpha$.
	\item {\bf Dispersal limited}: Each sample was stochastically colonized with a randomly chosen number of species, uniformly distributed between $S = 1$ and $S = S_{\rm tot} = 180$, and supplied with a single external resource, with $\kappa_1 = 200$ and $\kappa_\alpha = 0$ for all $\alpha \neq 1$.
\end{itemize}

\subsection{Synthetic data for human microbiome patterns}
To generate synthetic data for Figures \ref{fig:HMP}-\ref{fig:DOC}, we assumed a regional species pool of size $S_{\rm tot}=5000$, with $M=300$ possible resource types. Resources were grouped into $T =6$ classes of 50 resource types each, labeled A through F. Microbial species were grouped into 6 specialist ``families'' of 800 species, with each family specializing in one resource class as described above. The remaining 200 species were designated as generalists, with no bias towards any one resource class. The consumption parameters were set to $c_0 = 0$, $c_1 = 1$ and $\mu_c = 10$ as for the previous set of simulations. The metabolic matrix sparsity was set to $s=0.3$, to reflect the actual sparsity of the E. Coli metabolic network \cite{wagner2001small}, and the secretions were allocated with $f_s = f_w = 0.45$. The $m_i$ were sampled from a Gaussian distribution with mean 1 and standard deviation $0.01$. Each community was supplied with the same total incoming energy flux $\kappa = \sum_\alpha \kappa_\alpha = 1,000$. For each scenario, we simulated 900 independent communities, evenly partitioned among three ``body sites'' with different environmental characteristics. 


The rest of the parameters were varied to construct eight different scenarios. For $S = 2,500$ ({\bf strong dispersal limitation}) and $S = 4,900$ ({\bf weak dispersal limitation}), we made the following four combinations of species properties and environmental conditions:
\begin{itemize}
	\item {\bf Metabolically distinct simple environments}: In each of the three simulated body sites, one resource was chosen from each of two resource classes (A+B, C+D, and E+F). The relative flux levels for these two resources were chosen for each of the 300 communities in the site by randomly sampling a number $a$ from a uniform distribution over the interval $[0,1]$, and then letting $(1-a)\kappa$ be the flux of the first resource, with $a\kappa$ from the second. Taxonomic structure was incorporated by setting a high strength of specialization $q=0.9$.  
	\item {\bf Metabolically distinct complex environments}: In each of the three simulated body sites, all resources were supplied from two resource classes (A+B, C+D, and E+F), with flux levels from all other classes still set to zero. The relative flux levels for 100 resource types in each site were randomly sampled for each community, by independently sampling 100 numbers $\tilde{\kappa}_\alpha$ from a uniform distribution over the interval $[0,1]$, and then setting $\kappa_\alpha = \kappa\frac{\tilde{\kappa}_\alpha}{\sum_\beta \tilde{\kappa}_\beta}$. Taxonomic structure was incorporated by setting a high strength of specialization $q=0.9$. 
	\item {\bf No taxonomic structure}: Same as ``simple environments'' above, except that taxonomic structure was removed by setting $q = 0$.
	\item {\bf Metabolically overlapping complex environments}: Same as ``metabolically distinct complex environments'' except that the 300 resources are randomly partitioned into three sets of 100, and each body site is supplied with resources from a different set.
\end{itemize}

\subsection{Synthetic data for marine microbiome patterns}
The abundance distributions in Figure \ref{fig:AD} were generated directly from the ``Simple environments''  and ``No taxonomic structure'' simulations described above for the human microbiome patterns. 

The tests of modular community assembly in Figure \ref{fig:modular} were performed using the same setup as ``Simple environments'' in the human microbiome simulations, but with just two ``body sites'' (A+B and C+D), and three values of $a$ (0, 1 and 0.5). The {\rm unstructured metabolism} control was performed by setting $T=F=1$, assigning all 300 resources to the same resource class before sampling the metabolic and consumer preference matrices. 

\subsection{Relative abundance distributions and Fisher log series}
To create Figure \ref{fig:AD}, we first downloaded the 16S OTU table from the Tara Oceans companion website (\url{http://ocean-microbiome.embl.de/companion.html}) \cite{Sunagawa2015}. We performed 300 independent rarefactions to a constant read depth of 10,000. For each possible number of reads, from 1 through the maximum observed, we plotted the number of species assigned that number of reads (``population size''), averaged over all rarefactions. 

In many ecological settings, it has been observed that the number $s(n)$ of species with $n$ individuals in a sample of $N$ total individuals closely follows the Fisher log series \cite{fisher1943relation, magurran2005species}:
\begin{align}
s(n) = \frac{\alpha}{n} x^n
\label{eq:fish}
\end{align}
where the parameters $x$ and $\alpha$ are determined from $N$ and the total number of observed species $S$ through the following equations:
\begin{align}
S &= -\alpha \ln (1-x)\\
N &= \frac{\alpha x}{1-x}.
\end{align}
In the first panel of Figure \ref{fig:AD} we plot Equation (\ref{eq:fish}) using $N = 10,000$ (the read depth we manually imposed for the rarefaction) and $S$ equal to the number of OTU's with more than 5 reads assigned to them in the original dataset.  

For the simulation data in Figure \ref{fig:AD}, we had the further advantage of having access to multiple independent trials under statistically similar conditions. Instead of averaging over multiple rarefactions generated from the same underlying dataset, we averaged over single samples of $N = 1,000$ individuals taken from each of the 900 parallel communities. We plotted the Fisher log series with $N = 1,000$ and $S$ equal to the number of species with nonzero abundance.

We also plotted the distributions obtained when the true relative proportions of $S$ species are generated by a simple null model, in which the invasion fitness (low-density growth rate) of each species is sampled from a Gaussian distribution. Species with negative values go extinct, and those with positive values end up with population sizes proportional to the invasion fitness. The resulting relative abundances of species in an infinitely large community follow a truncated Gaussian distribution. This distribution is determined by a single parameter, up to an overall scale that is irrelevant for the purposes of the current analysis. The parameter is inferred from the simulation results by matching the fraction of species initially present in the community that survive to equilibrium. In the figure, the green curves come from first sampling $S$ true species abundances from this distribution, then sampling $N = 1,000$ individuals from the resulting population, and averaging the results over 10,000 independent iterations. 

\subsection{Computation of overlap and dissimilarity}
Here we summarize the definitions of the dissimilarity $D(\{N_i^\mu\},\{N_j^\nu\})$ and overlap $O(\{N_i^\mu\},\{N_j^\nu\})$ between two sets of population size measurements, as given in \cite{bashan2016universality}. Here, $\mu = 1, 2,\dots C$ is an index labeling the sample from which the measurement was taken (as is $\nu$). In order to define these two quantities, we must first introduce some notation concerning the shared species. We let $\mathbf{S}^\dagger$ represent the set of species that are present in both communities, and denote the total number of species in this set by $S^\dagger$. We also define two types of normalized abundances:
\begin{align}
\tilde{N}_i &= \frac{N_i}{\sum_{j} N_j}\\
\hat{N}_i &= \frac{N_i}{\sum_{j\in\mathbf{S}^\dagger} N_j}
\end{align}
where the second quantity is normalized only by the set of species that is shared with the other community in the pair. We also define the average composition over the shared species:
\begin{align}
m_i = \frac{1}{2} (\hat{N}_i^\mu + \hat{N}_i^\nu).
\end{align}
Using these definitions, we can finally write
\begin{align}
D(\{N_i^\mu\},\{N_j^\nu\}) &= \sqrt{\frac{1}{2}\sum_{i\in\mathbf{S}^\dagger}\left(\hat{N}_i^\mu \ln \frac{\hat{N}_i^\mu}{m_i} + \hat{N}_i^\nu \ln \frac{\hat{N}_i^\nu}{m_i}\right)}\\
O(\{N_i^\mu\},\{N_j^\nu\}) &= \frac{1}{2}\sum_{i \in \mathbf{S}^\dagger}(\tilde{N}_i^\mu + \tilde{N}_i^\nu).
\end{align}
The first equation is simply the square root of the Jensen-Shannon divergence between the relative abundances of the overlapping species, and the second measures the relative abundance of the species in the overlapping set, averaged over the two communities.

\subsection{Computation of effective Lotka-Volterra parameters}
The distinction between ``host-specific'' vs. ``universal'' population dynamics is most clearly defined in terms of a closed set of equations for the dynamics of the population sizes, with environmental factors treated implicitly \cite{bashan2016universality}. We can transform the MiCRM into a model of this form by examining the regime where the resource dynamics are ``fast'' compared to the timescale for changes in population sizes. We can then simplify the form of the resulting model by performing a Taylor expansion of the growth rate around the equilibrium population sizes $\bar{N}_i$, resulting in generalized Lotka-Volterra equations parameterized by a set of carrying capacities and interaction coefficients.

We start by writing the full dynamical equations (\ref{eq:MiCRM}) in a more compact form:
\begin{align}
\frac{dN_i}{dt} &= g_i N_i\left[ \sum_\alpha \tilde{w}_\alpha c_{i\alpha} R_\alpha - m_i\right]\\
\frac{dR_\alpha}{dt} &= \kappa_\alpha - \tau^{-1} R_\alpha - \sum_{j\beta}Q_{\alpha\beta}N_j c_{j\beta} R_\beta
\end{align}
with
\begin{align}
\tilde{w}_\alpha &= w_\alpha (1-l_\alpha)\\
Q_{\alpha\beta} &= \delta_{\alpha\beta} - D_{\alpha\beta}\frac{w_\beta}{w_\alpha}l_\beta.
\end{align}
We now invoke the ``fast resource equilibration'' assumption to set $dR_\alpha/dt = 0$, and solve for the resource concentrations $\bar{R}_\alpha$ as functions of the set of population sizes $\{N_j\}$. Inserting this result back into the equations for the dynamics of $N_i$, we have:
\begin{align}
\frac{dN_i}{dt} &= g_i N_i \left[ \sum_\alpha \tilde{w}_\alpha c_{i\alpha} \bar{R}_\alpha(\{N_j\}) - m_i\right].
\end{align}
To obtain the local Lotka-Volterra coefficients, we perform a Taylor expansion of the term in brackets around $N_j = \bar{N}_j$, up to first order in the distance from equilibrium:
\begin{align}
\frac{dN_i}{dt} &= g_i N_i \left[ \sum_{\alpha j} \tilde{w}_\alpha c_{i\alpha} \frac{\partial \bar{R}}{\partial N_j}(N_j - \bar{N}_j) - m_i + \mathcal{O}((N_j-\bar{N}_j)^2)\right]\\
&= \frac{r_i}{K_i} N_i \left[K_i - N_i - \sum_{j\neq i} \alpha_{ij} N_j + \mathcal{O}((N_j-\bar{N}_j)^2)\right]
\end{align}
where
\begin{align}
\alpha_{ij} &= -\frac{\sum_{\alpha} \tilde{w}_\alpha c_{i\alpha} \frac{\partial \bar{R}}{\partial N_j}}{\sum_{\beta} \tilde{w}_\beta c_{i\beta} \frac{\partial \bar{R}}{\partial N_i}}\\
K_i &= \sum_j \alpha_{ij} \bar{N}_j - m_i\\
r_i &= K_i g_i \sum_{\alpha} \tilde{w}_\alpha c_{i\alpha} \frac{\partial \bar{R}}{\partial N_i}.
\end{align}
We can compute the derivatives of $\bar{R}$ through implicit differentiation to obtain
\begin{align}
\sum_\beta A_{\alpha\beta} \frac{\partial \bar{R}_\beta}{\partial N_j} &= - \sum_\beta Q_{\alpha\beta} c_{j\beta} \bar{R}_\beta
\end{align}
with
\begin{align}
A_{\alpha\beta} = \tau^{-1} \delta_{\alpha\beta} + Q_{\alpha\beta} \sum_i \bar{N}_i c_{i\beta}.
\end{align}
Thus we find
\begin{align}
\frac{\partial \bar{R}_\alpha}{\partial N_j} &= -\sum_{\beta\gamma} (A^{-1})_{\alpha\beta} Q_{\beta\gamma} c_{j\gamma} \bar{R}_\gamma
\end{align}
and conclude
\begin{align}
\alpha_{ij} = \frac{\sum_{\alpha\beta} c_{i\alpha} W_{\alpha\beta} c_{j\beta}}{\sum_{\gamma\delta} c_{i\gamma} W_{\gamma\delta} c_{i\delta}}
\end{align}
where 
\begin{align}
W_{\alpha\beta} = \sum_\gamma \tilde{w}_\alpha (A^{-1})_{\alpha\gamma} Q_{\gamma\beta} \bar{R}_\beta.
\end{align}

Since all the parameters and equilibrium abundances are known in the simulations, this set of equations allows us to compute for each community $\mu$ ($\mu = 1, 2 \dots 300$) the bare growth rates $r_i^{\mu}$, the interactions $\alpha_{ij}^\mu$ and the carrying capacities $K_i^\mu$. The scatter plot in the bottom right panel of Figure \ref{fig:DOC} shows the normalized RMS variability in the carrying capacities for each pair of samples $\mu$ and $\nu$, computed as:
\begin{align}
{\rm variability}_{\mu\nu} = \frac{\sqrt{\frac{1}{S^\dagger}\sum_{i\in \mathbf{S}^\dagger} (K_i^\mu - K_i^\nu)^2}}{\frac{1}{2S^\dagger}\sum_{j\in \mathbf{S}^\dagger} (K_j^\mu + K_j^\nu)}
\end{align}
where $S^\dagger$ is the number of surviving species shared between the two communities, and $\mathbf{S}^\dagger$ is the set of indices of the shared species.

\section*{Acknowledgments}
This work was supported by NIH NIGMS grant 1R35GM119461 and Simons Investigator in the Mathematical Modeling of Living Systems (MMLS) award to PM. Computational work was performed on the Shared Computing Cluster which is administered by Boston University Research Computing Services. 

\section*{Competing Interests}
The authors declare no competing interests.


\section*{References}
\begin{thebibliography}{10}
	\expandafter\ifx\csname url\endcsname\relax
	\def\url#1{\texttt{#1}}\fi
	\expandafter\ifx\csname urlprefix\endcsname\relax\def\urlprefix{URL }\fi
	\providecommand{\bibinfo}[2]{#2}
	\providecommand{\eprint}[2][]{\url{#2}}
	
	\bibitem{EMP}
	\bibinfo{author}{Thompson, L.~R.} \emph{et~al.}
	\newblock \bibinfo{title}{{A communal catalogue reveals Earth's multiscale
			microbial diversity}}.
	\newblock \emph{\bibinfo{journal}{Nature}} \textbf{\bibinfo{volume}{551}},
	\bibinfo{pages}{457} (\bibinfo{year}{2017}).
	
	\bibitem{HMP}
	\bibinfo{author}{Huttenhower, C.} \emph{et~al.}
	\newblock \bibinfo{title}{Structure, function and diversity of the healthy
		human microbiome}.
	\newblock \emph{\bibinfo{journal}{Nature}} \textbf{\bibinfo{volume}{486}},
	\bibinfo{pages}{207} (\bibinfo{year}{2012}).
	
	\bibitem{qin2010human}
	\bibinfo{author}{Qin, J.} \emph{et~al.}
	\newblock \bibinfo{title}{A human gut microbial gene catalogue established by
		metagenomic sequencing}.
	\newblock \emph{\bibinfo{journal}{Nature}} \textbf{\bibinfo{volume}{464}},
	\bibinfo{pages}{59} (\bibinfo{year}{2010}).
	
	\bibitem{levin1992problem}
	\bibinfo{author}{Levin, S.~A.}
	\newblock \bibinfo{title}{{The problem of pattern and scale in ecology: the
			Robert H. MacArthur award lecture}}.
	\newblock \emph{\bibinfo{journal}{Ecology}} \textbf{\bibinfo{volume}{73}},
	\bibinfo{pages}{1943--1967} (\bibinfo{year}{1992}).
	
	\bibitem{hart2019uncovering}
	\bibinfo{author}{Hart, S.~F.} \emph{et~al.}
	\newblock \bibinfo{title}{Uncovering and resolving challenges of quantitative
		modeling in a simplified community of interacting cells}.
	\newblock \emph{\bibinfo{journal}{PLoS biology}} \textbf{\bibinfo{volume}{17}},
	\bibinfo{pages}{e3000135} (\bibinfo{year}{2019}).
	
	\bibitem{Wigner1955}
	\bibinfo{author}{Wigner, E.~P.}
	\newblock \bibinfo{title}{Characteristic vectors of bordered matrices with
		infinite dimensions}.
	\newblock \emph{\bibinfo{journal}{Annals of Mathematics (ser. 2)}}
	\textbf{\bibinfo{volume}{62}}, \bibinfo{pages}{548} (\bibinfo{year}{1955}).
	
	\bibitem{May1972}
	\bibinfo{author}{May, R.}
	\newblock \bibinfo{title}{{Will a Large Complex System be Stable?}}
	\newblock \emph{\bibinfo{journal}{Nature}} \textbf{\bibinfo{volume}{238}},
	\bibinfo{pages}{413} (\bibinfo{year}{1972}).
	
	\bibitem{Goldford2018}
	\bibinfo{author}{Goldford, J.~E.} \emph{et~al.}
	\newblock \bibinfo{title}{Emergent simplicity in microbial community assembly}.
	\newblock \emph{\bibinfo{journal}{Science}} \textbf{\bibinfo{volume}{361}},
	\bibinfo{pages}{469} (\bibinfo{year}{2018}).
	
	\bibitem{marsland2018available}
	\bibinfo{author}{Marsland~III, R.} \emph{et~al.}
	\newblock \bibinfo{title}{Available energy fluxes drive a transition in the
		diversity, stability, and functional structure of microbial communities}.
	\newblock \emph{\bibinfo{journal}{PLOS Computational Biology}}
	\textbf{\bibinfo{volume}{15}}, \bibinfo{pages}{e1006793}
	(\bibinfo{year}{2019}).
	
	\bibitem{marsland2019community}
	\bibinfo{author}{Marsland~III, R.}, \bibinfo{author}{Cui, W.},
	\bibinfo{author}{Golford, J.} \& \bibinfo{author}{Mehta, P.}
	\newblock \bibinfo{title}{{The Community Simulator: A Python package for
			microbial ecology}} \bibinfo{pages}{arXiv:1904.09367} (\bibinfo{year}{2019}).
	
	\bibitem{MacArthur1970}
	\bibinfo{author}{MacArthur, R.}
	\newblock \bibinfo{title}{{Species Packing and Competitive Equilibrium for Many
			Species}}.
	\newblock \emph{\bibinfo{journal}{Theoretical Population Biology}}
	\textbf{\bibinfo{volume}{1}}, \bibinfo{pages}{1} (\bibinfo{year}{1970}).
	
	\bibitem{Harcombe2014}
	\bibinfo{author}{Harcombe, W.~R.} \emph{et~al.}
	\newblock \bibinfo{title}{Metabolic resource allocation in individual microbes
		determines ecosystem interactions and spatial dynamics}.
	\newblock \emph{\bibinfo{journal}{Cell Reports}} \textbf{\bibinfo{volume}{7}},
	\bibinfo{pages}{1104} (\bibinfo{year}{2014}).
	
	\bibitem{Zomorrodi2016}
	\bibinfo{author}{Zomorrodi, A.~R.} \& \bibinfo{author}{Segr\`{e}, D.}
	\newblock \bibinfo{title}{Synthetic ecology of microbes: Mathematical models
		and applications}.
	\newblock \emph{\bibinfo{journal}{J. Mol. Biol.}}
	\textbf{\bibinfo{volume}{428}}, \bibinfo{pages}{837} (\bibinfo{year}{2016}).
	
	\bibitem{pacheco2019costless}
	\bibinfo{author}{Pacheco, A.~R.}, \bibinfo{author}{Moel, M.} \&
	\bibinfo{author}{Segr\`{e}, D.}
	\newblock \bibinfo{title}{Costless metabolic secretions as drivers of
		interspecies interactions in microbial ecosystems}.
	\newblock \emph{\bibinfo{journal}{Nature Communications}}
	\textbf{\bibinfo{volume}{10}}, \bibinfo{pages}{103} (\bibinfo{year}{2019}).
	
	\bibitem{leibold2004metacommunity}
	\bibinfo{author}{Leibold, M.~A.} \emph{et~al.}
	\newblock \bibinfo{title}{The metacommunity concept: a framework for
		multi-scale community ecology}.
	\newblock \emph{\bibinfo{journal}{Ecology Letters}}
	\textbf{\bibinfo{volume}{7}}, \bibinfo{pages}{601} (\bibinfo{year}{2004}).
	
	\bibitem{vellend2016theory}
	\bibinfo{author}{Vellend, M.}
	\newblock \emph{\bibinfo{title}{{The Theory of Ecological Communities
				(MPB-57)}}}, vol.~\bibinfo{volume}{75} (\bibinfo{publisher}{Princeton
		University Press}, \bibinfo{year}{2016}).
	
	\bibitem{hillerislambers2012rethinking}
	\bibinfo{author}{HilleRisLambers, J.}, \bibinfo{author}{Adler, P.},
	\bibinfo{author}{Harpole, W.}, \bibinfo{author}{Levine, J.} \&
	\bibinfo{author}{Mayfield, M.}
	\newblock \bibinfo{title}{Rethinking community assembly through the lens of
		coexistence theory}.
	\newblock \emph{\bibinfo{journal}{Annual Review of Ecology, Evolution, and
			Systematics}} \textbf{\bibinfo{volume}{43}} (\bibinfo{year}{2012}).
	
	\bibitem{dini2018embracing}
	\bibinfo{author}{Dini-Andreote, F.} \& \bibinfo{author}{Raaijmakers, J.~M.}
	\newblock \bibinfo{title}{Embracing community ecology in plant microbiome
		research}.
	\newblock \emph{\bibinfo{journal}{Trends in Plant Science}}
	\textbf{\bibinfo{volume}{23}}, \bibinfo{pages}{467--469}
	(\bibinfo{year}{2018}).
	
	\bibitem{shurin2000dispersal}
	\bibinfo{author}{Shurin, J.~B.}
	\newblock \bibinfo{title}{Dispersal limitation, invasion resistance, and the
		structure of pond zooplankton communities}.
	\newblock \emph{\bibinfo{journal}{Ecology}} \textbf{\bibinfo{volume}{81}},
	\bibinfo{pages}{3074} (\bibinfo{year}{2000}).
	
	\bibitem{Sunagawa2015}
	\bibinfo{author}{Sunagawa, S.} \emph{et~al.}
	\newblock \bibinfo{title}{{Structure and function of the global ocean
			microbiome}}.
	\newblock \emph{\bibinfo{journal}{Science}} \textbf{\bibinfo{volume}{348}},
	\bibinfo{pages}{1261359} (\bibinfo{year}{2015}).
	
	\bibitem{enke2018modular}
	\bibinfo{author}{Enke, T.~N.} \emph{et~al.}
	\newblock \bibinfo{title}{Modular assembly of polysaccharide-degrading
		microbial communities in the ocean}.
	\newblock \emph{\bibinfo{journal}{Current Biology}}
	\textbf{\bibinfo{volume}{29}}, \bibinfo{pages}{1528} (\bibinfo{year}{2019}).
	
	\bibitem{mehta2018constrained}
	\bibinfo{author}{Mehta, P.}, \bibinfo{author}{Cui, W.}, \bibinfo{author}{Wang,
		C.-H.} \& \bibinfo{author}{Marsland~III, R.}
	\newblock \bibinfo{title}{Constrained optimization as ecological dynamics with
		applications to random quadratic programming in high dimensions}.
	\newblock \emph{\bibinfo{journal}{Physical Review E}}
	\textbf{\bibinfo{volume}{99}}, \bibinfo{pages}{052111}
	(\bibinfo{year}{2018}).
	
	\bibitem{Marsland2019a}
	\bibinfo{author}{Marsland~III, R.}, \bibinfo{author}{Cui, W.} \&
	\bibinfo{author}{Mehta, P.}
	\newblock \bibinfo{title}{The minimum environmental perturbation principle: A
		new perspective on niche theory} \bibinfo{pages}{arXiv:1901.09673}
	(\bibinfo{year}{2019}).
	
	\bibitem{hoehler2013microbial}
	\bibinfo{author}{Hoehler, T.~M.} \& \bibinfo{author}{J{\o}rgensen, B.~B.}
	\newblock \bibinfo{title}{Microbial life under extreme energy limitation}.
	\newblock \emph{\bibinfo{journal}{Nature Reviews Microbiology}}
	\textbf{\bibinfo{volume}{11}}, \bibinfo{pages}{83} (\bibinfo{year}{2013}).
	
	\bibitem{fisher1943relation}
	\bibinfo{author}{Fisher, R.~A.}, \bibinfo{author}{Corbet, A.~S.} \&
	\bibinfo{author}{Williams, C.~B.}
	\newblock \bibinfo{title}{The relation between the number of species and the
		number of individuals in a random sample of an animal population}.
	\newblock \emph{\bibinfo{journal}{The Journal of Animal Ecology}}
	\bibinfo{pages}{42--58} (\bibinfo{year}{1943}).
	
	\bibitem{magurran2005species}
	\bibinfo{author}{Magurran, A.~E.}
	\newblock \bibinfo{title}{Species abundance distributions: pattern or process?}
	\newblock \emph{\bibinfo{journal}{Functional Ecology}}
	\textbf{\bibinfo{volume}{19}}, \bibinfo{pages}{177--181}
	(\bibinfo{year}{2005}).
	
	\bibitem{Advani2018}
	\bibinfo{author}{Advani, M.}, \bibinfo{author}{Bunin, G.} \&
	\bibinfo{author}{Mehta, P.}
	\newblock \bibinfo{title}{{Statistical physics of community ecology: a cavity
			solution to MacArthur's consumer resource model}}.
	\newblock \emph{\bibinfo{journal}{Journal of Statistical Mechanics}}
	\textbf{\bibinfo{volume}{2018}}, \bibinfo{pages}{033406}
	(\bibinfo{year}{2018}).
	
	\bibitem{hubbell_unified_2001}
	\bibinfo{author}{Hubbell, S.~P.}
	\newblock \emph{\bibinfo{title}{The Unified Neutral Theory of Biodiversity and
			Biogeography ({MPB-32)}}} (\bibinfo{publisher}{Princeton University Press},
	\bibinfo{year}{2001}).
	
	\bibitem{pearce2019stabilization}
	\bibinfo{author}{Pearce, M.~T.}, \bibinfo{author}{Agarwala, A.} \&
	\bibinfo{author}{Fisher, D.~S.}
	\newblock \bibinfo{title}{Stabilization of extensive fine-scale diversity by
		spatio-temporal chaos}.
	\newblock \emph{\bibinfo{journal}{bioRxiv}} \bibinfo{pages}{736215}
	(\bibinfo{year}{2019}).
	
	\bibitem{costea2018enterotypes}
	\bibinfo{author}{Costea, P.~I.} \emph{et~al.}
	\newblock \bibinfo{title}{Enterotypes in the landscape of gut microbial
		community composition}.
	\newblock \emph{\bibinfo{journal}{Nature Microbiology}}
	\textbf{\bibinfo{volume}{3}}, \bibinfo{pages}{8} (\bibinfo{year}{2018}).
	
	\bibitem{gorvitovskaia2016interpreting}
	\bibinfo{author}{Gorvitovskaia, A.}, \bibinfo{author}{Holmes, S.~P.} \&
	\bibinfo{author}{Huse, S.~M.}
	\newblock \bibinfo{title}{Interpreting prevotella and bacteroides as biomarkers
		of diet and lifestyle}.
	\newblock \emph{\bibinfo{journal}{Microbiome}} \textbf{\bibinfo{volume}{4}},
	\bibinfo{pages}{15} (\bibinfo{year}{2016}).
	
	\bibitem{Arumugam2011}
	\bibinfo{author}{Arumugam, M.} \emph{et~al.}
	\newblock \bibinfo{title}{Enterotypes of the human gut microbiome}.
	\newblock \emph{\bibinfo{journal}{Nature}} \textbf{\bibinfo{volume}{473}},
	\bibinfo{pages}{174} (\bibinfo{year}{2011}).
	
	\bibitem{bashan2016universality}
	\bibinfo{author}{Bashan, A.} \emph{et~al.}
	\newblock \bibinfo{title}{Universality of human microbial dynamics}.
	\newblock \emph{\bibinfo{journal}{Nature}} \textbf{\bibinfo{volume}{534}},
	\bibinfo{pages}{259} (\bibinfo{year}{2016}).
	
	\bibitem{kalyuzhny2017dissimilarity}
	\bibinfo{author}{Kalyuzhny, M.} \& \bibinfo{author}{Shnerb, N.~M.}
	\newblock \bibinfo{title}{Dissimilarity-overlap analysis of community dynamics:
		Opportunities and pitfalls}.
	\newblock \emph{\bibinfo{journal}{Methods in Ecology and Evolution}}
	\textbf{\bibinfo{volume}{8}}, \bibinfo{pages}{1764} (\bibinfo{year}{2017}).
	
	\bibitem{barbier2018generic}
	\bibinfo{author}{Barbier, M.}, \bibinfo{author}{Arnoldi, J.-F.},
	\bibinfo{author}{Bunin, G.} \& \bibinfo{author}{Loreau, M.}
	\newblock \bibinfo{title}{Generic assembly patterns in complex ecological
		communities}.
	\newblock \emph{\bibinfo{journal}{Proceedings of the National Academy of
			Sciences}} \textbf{\bibinfo{volume}{115}}, \bibinfo{pages}{2156--2161}
	(\bibinfo{year}{2018}).
	
	\bibitem{cui2019diverse}
	\bibinfo{author}{Cui, W.}, \bibinfo{author}{Marsland~III, R.} \&
	\bibinfo{author}{Mehta, P.}
	\newblock \bibinfo{title}{Diverse communities behave like typical random
		ecosystems} \bibinfo{pages}{arXiv:1904.02610} (\bibinfo{year}{2019}).
	
	\bibitem{csillery2010approximate}
	\bibinfo{author}{Csill{\'e}ry, K.}, \bibinfo{author}{Blum, M.~G.},
	\bibinfo{author}{Gaggiotti, O.~E.} \& \bibinfo{author}{Fran{\c{c}}ois, O.}
	\newblock \bibinfo{title}{{Approximate Bayesian computation (ABC) in
			practice}}.
	\newblock \emph{\bibinfo{journal}{Trends in Ecology \& Evolution}}
	\textbf{\bibinfo{volume}{25}}, \bibinfo{pages}{410--418}
	(\bibinfo{year}{2010}).
	
	\bibitem{wagner2001small}
	\bibinfo{author}{Wagner, A.} \& \bibinfo{author}{Fell, D.~A.}
	\newblock \bibinfo{title}{The small world inside large metabolic networks}.
	\newblock \emph{\bibinfo{journal}{Proceedings of the Royal Society of London B:
			Biological Sciences}} \textbf{\bibinfo{volume}{268}}, \bibinfo{pages}{1803}
	(\bibinfo{year}{2001}).
	
\end{thebibliography}

\end{document}